\documentclass[preprint]{aastex}

 \usepackage{epsfig}
 \usepackage{longtable}
 \usepackage{graphicx}
 \usepackage{natbib}

 \usepackage[colorlinks=true,urlcolor=black,linkcolor=black,citecolor=black]{hyperref} 

 \def\etal{{\it et al.}\thinspace}
 \def\ie{{\it i.e.,}\thinspace}
 \def\eg{{\it e.g.,}\thinspace}
 \newcommand{\revision}[1]{\textbf{#1}} 

 \usepackage{xcolor}
 \newcounter{attnctr} 

\begin{document}
 \title{Polarized quasiperiodic structures in pulsar radio emission reflect temporal modulations of non-stationary plasma flow}
 \shorttitle{Polarized quasiperiodic structures in pulsar radio emission}
 \shortauthors{Mitra, Arjunwadkar \& Rankin}

 \author{Dipanjan Mitra} 
 \affil{National Centre for Radio Astrophysics, TIFR, Pune 411007, India; 
 email: dmitra@ncra.tifr.res.in}
 \and
 \author{Mihir Arjunwadkar} 
 \affil{Centre for Modeling and Simulation, Savitribai Phule Pune University, Pune 411007, India; 
 email: mihir@cms.unipune.ac.in}
 \and
 \author{Joanna M. Rankin}
 \affil{University of Vermont, Burlington, VT 05405 USA; 
 email: Joanna.Rankin@uvm.edu}

\begin{abstract}
Bright single pulses of many radio pulsars show rapid intensity
fluctuations (called microstructure) when observed with time
resolutions of tens of microseconds.  Here, we report an analysis of
Arecibo 59.5 $\mu$sec-resolution polarimetric observations of 11
P-band and 32 L-band pulsars with periods ranging from 150 msec to 3.7
sec.  These higher frequency observations forms the most
reliable basis for detailed microstructure studies.  Close inspection
of individual pulses reveals that most pulses exhibit
quasiperiodicities with a well-defined periodicity timescale
($P_{\mu}$).  While we find some pulses with deeply modulating
microstructure, most pulses show low-amplitude modulations on top of
broad smooth subpulses features, thereby making it difficult to infer
periodicities.  We have developed a method for such low-amplitude
fluctuations wherein a smooth subpulse envelope is subtracted from
each de-noised subpulse; the fluctuating portion of each subpulse is
then used to estimate $P_{\mu}$ via autocorrelation analysis.  We find
that the microstructure timescale $P_{\mu}$ is common across all
Stokes parameters of polarized pulsar signals.  Moreover, no clear
signature of curvature radiation in vacuum in highly resolved
microstructures was found.  Our analysis further shows strong
correlation between $P_\mu$ and the pulsar period $P$.  We discuss
implications of this result in terms of a coherent radiation model
wherein radio emission arises due to formation and acceleration of
electron-positron pairs in an inner vacuum gap over magnetic polar
cap, and a subpulse corresponds to a series of 
non-stationary sparking discharges.
We argue that in this model, $P_{\mu}$ reflects the temporal
modulation of non-stationary plasma flow.
\end{abstract}


\section{Introduction}
\label{intro}
Soon after the discovery of pulsars, \citet{1968Natur.218.1122C}
noticed sub-millisecond (\ie hundreds of microseconds) intensity
structures in single pulses of PSR B0950+08.  Subsequently, several
high-time-resolution studies revealed rapid intensity variations
superposed on the subpulses of several other pulsars.  This
phenomenon, usually known as pulsar microstructure, is currently
thought to be associated closely with mechanisms of magnetospheric
radio emission.  Observationally, microstructures are found to occur
simultaneously over a wide band (\eg \citealt{1975ApJ...201..425R})
and in pulsars with rotation periods down to 89 msec
\citep{2002MNRAS.334..523K}, providing strong empirical evidence in
favor of magnetospheric emission.  The high brightness temperatures
observed in pulsars rule out any thermal plasma processes, and
indicate that the underlying mechanism must be a coherent plasma
process.  However, a coherent radiation mechanism that provides a
physical interpretation of microstructure behavior is yet to be
identified.

Broadly, two appealing scenarios have been considered to explain
microstructure emission, commonly known as angular beaming or temporal
modulation.  The angular beaming scheme is a geometrical phenomenon
wherein intensity fluctuations result from an angular radiation
pattern (mimicking microstructure width and periodicity) in the
transverse direction.  As the pulsar rotates, an observer then samples
this emission structure.  On the other hand, the temporal modulation
scheme entails intensity fluctuation of propagating waves in the
magnetosphere that finally reach the observer.

The existing microstructure studies fall into two broad categories:
Studies with very high time resolution ($\sim$100 nanosec) which are
needed to study narrow microstructures; and those with a coarser
resolution ($\gtrsim$10 $\mu$sec) that are useful for exploring the
properties of broad microstructures.  Amongst the former is the
well-known detection of ultrafast 2-nanosecond temporal structures in
the Crab pulsar (\citealt{2003Natur.422..141H}; see also
\citealt{2010A&A...524A..60J}).  Intensity fluctuations with time
scales as small as 2.5 $\mu$sec have also been reported in PSR
B1133+16 by \citet{1982ApJ...254L..35B}, and
\citet{2002A&A...396..171P} find micropulses with widths between 2 and
7 $\mu$sec in pulsars B0950+08, B1133+16 and B1929+10.  Most of the
other well-known microstructure studies were geared to study the broad
microstructures.  These have shown that subpulses often show
quasiperiodic intensity variations with periods of hundreds of
microseconds (\eg \citealt{1971ApJ...169..487H};
\citealt{1976ApJ...208L..43B}; \citealt{1981SvA....25..442S};
\citealt{1983SvA....27..169S};
\citealt{1990AJ....100.1882C}). \citet{1979AuJPh..32....9C} and
\citet{2002MNRAS.334..523K} further show that the measured temporal
width ($t_{\mu}$) of a micropulse is related linearly to the rotation
period $P$ approximately as $t_{\mu} \mbox{ (msec)} \sim 10^{-3} P
\mbox{ (sec)}$ and $t_{\mu} \mbox{ (msec)} \sim 0.7 \times 10^{-3} P
\mbox{ (sec)}$ respectively.  Polarized microstructure has been
investigated for a few pulsars (\citealt{1978A&A....64...27F};
\citealt{1979AuJPh..32....9C}), and only for the Vela pulsar B0833--45
does some quantitative description exist \citep{2002MNRAS.334..523K}.
Further observations, analyses, and phenomenological assessments are
needed to establish how microstructure is connected physically with
the mechanisms of pulsar emission.

One of the difficulties in using the existing microstructure studies
to amalgamate with pulsar emission theories is that they represent
different resolutions, frequencies of observation, and analysis
methods.  Hence, in the remainder of our work, our main effort is to
(1) carry out a systematic, single-frequency, single instrument
(Arecibo) study of the properties of broad polarized microstructure;
and (2) study the physical implications of the microstructure
characteristics in the context of currently-favored coherent radio
emission theories.

\section{Observations}
\label{sec1}
High-time-resolution single-pulse polarimetric observations were
carried out using the Arecibo telescope for 32 pulsars in the
frequency range of about 1.2 -- 1.6 GHz (L--band), and 11 pulsars at
around 327 MHz (P--band). For these observations, we used the
Observatory's Mock spectrometers in the ``single-pixel'' polarimetry
mode (see \url{http://www.naic.edu/$\sim$astro/mock.shtml}).  These
Fourier-transform-based instruments are capable of sampling at
$\sim$59.5 $\mu$sec rate and producing Stokes spectra in a filterbank
format.  The details of the observations are given in
Table~\ref{tab1}.  This moderate time resolution was chosen both as
suitable for investigating broad microstructure and maximizing the
number of objects that could be so observed.  In the remainder of this
paper, we will use the word microstructure to imply broad
microstructure unless specified otherwise.  Heretofore, only some 14
well-known bright pulsars have been subjected to microstructure
studies (see Table 2 of \citealt{2002MNRAS.334..523K} and the Vela
Pulsar), out of which 11 are in the Arecibo declination range.  Our
sample included 10 of these previously studied objects (except PSR
B0540+23) at L band as well as 22 others that are now accessible owing
to both the Gregorian feed system and capacious Mock back-ends. At P
band, there is an overlap with 6 pulsars; however, for most pulsars at
this frequency (apart from B1929+10), either the dispersion or the
scattering timescale is comparable with the sampling rate and hence
they are not suitable for microstructure study.  Nonetheless, given
the novelty of this 59.5-$\mu$sec polarimetry with excellent signal to
noise (S/N) ratio, we do display their properties in the online
supplementary material. The high-time-resolution average polarization
profiles of 24 pulsars at L band which were eventually used for
microstructure study (see Sec.\ \ref{sec2}) are shown in
Sec.\ \ref{a_avprof}, while the rest are available in the online
supplementary material.  The pulsar position angles are rotated to
infinite frequency using the rotation measure values given in
Table~\ref{tab1}~and, in principle, are the absolute position
angles. We note that the absolute position angles of our observations
at L band are more robust and the position angles appear to be in good
agreement with earlier published results (\eg
\citealt{1981A&AS...46..421M}; \citealt{2005MNRAS.364.1397J};
\citealt{2007ApJ...664..443R}).  In P band, the bandwidths used for
our observations were not large enough to calculate accurate rotation
measures, and hence an error in position angles of about 1 rad
m$^{-2}$ (which can result due to the ionosphere) cannot be ruled
out \footnote{ Note, that the P band observations will therefore not
  be included in the analyses below or discussed therein.}.

The average polarization properties of the pulsars in our study agree
with lower-resolution observations reported in several other studies
(\eg \citealt{2001ApJ...553..341E}).  None of these average profiles
show any obvious signatures of microstructure, and are consistent with
the view that, at any given longitude, the averaging process washes
out any such structure that can be seen in individual pulses. The
pulse-phase-averaged percentage linear and circular polarization
values over bins for signal to noise ratio greater than 3 (where the
noise was estimated as off pulse rms values for linear and circular
polarization respectively) for all our observations are given in
Table~\ref{tab1}.  The pulse window or phase region over which the
averaging was done corresponds to outer edges of the pulse profile at
three times the off pulse rms level of the total intensity In
Table~\ref{tab1}, we have not included error estimates on \%$L$ and
\%$V$.  RMS values could be computed using standard methods of error
propagation [see, e.g., \citet{2015arXiv150103312N}] assuming Gaussian
distributions.  However, rigorous error estimates are best-quantified
in terms of valid confidence intervals, and we believe that such
computation for \%$L \propto L/I$ and \%$V \propto V/I$ may not be
straightforward.  While the noise in the measured Stokes parameters
$I,Q,U,V$ is Gaussian, the distribution of the ratio of two Gaussian
quantities (i.e., $V/I$) is not Gaussian (e.g., the ratio of two
mean-zero Gaussian quantities has a Cauchy or Lorentz distribution;
and the distribution of $L=\sqrt{Q^2+U^2}$ is skewed and non-Gaussian
such that the ratio \%$L$ is non-Gaussian as well).  Given the
non-Gaussianity of \%$L$ and \%$V$, characterizing their errors as an
RMS serves no purpose.  Stokes parameter measurements are also subject
to systematic errors due to cross-coupling in the feed system.  The
crossed-linear P and L-band systems at Arecibo have isolations in
voltage of some 25 db or better (see Arecibo technical memos in
http://www.naic.edu/$\sim$astro/aotms/ and
http://www.naic.edu/$\sim$phil/stokes/stokescal.html), or in power of
about 0.5\%.  Our observations correct for the instrumental gains and
phase using a linearly polarized calibration signal, but no correction
was made for the cross-coupling.  We except this latter to introduce a
maximum mutual error of 5\% in $L$ and $V$ (e.g., $\Delta L \sim 0.05
V$ and $\Delta V \sim 0.05 L$).

Single pulses, however, exhibit highly complex structures, with
intensity fluctuations over timescales much smaller than the widths of
their subpulses.  Some obvious conclusions can be drawn from a visual
inspection of individual pulses.  For example, in several pulsars, we
find that there are occasional bright pulses where the subpulses show
a highly modulated ringing structure with all the Stokes parameters
fluctuating with the same periodicity. Several such examples are shown
in Fig.~\ref{fig1}.  However, almost all subpulses show low-amplitude
quasiperiodic intensity variations superposed on a broad smooth
envelope.  In the subsequent sections, we will develop and apply a
method of extracting microstructure periodicities for both
low-amplitude and highly-modulated subpulses.  The microstructure
emission is also seen in both the orthogonal polarization emission
modes (an example is shown in Fig.~\ref{fig2}).

\section{Estimating the microstructure timescale for an individual pulse}
\label{sec2}

The focus of our analysis is on extracting microstructure timescales
from time-series data for individual pulses.  Timescales are
traditionally identified as minima, maxima, or slope changes in the
autocorrelation function (ACF) for a pulse.  We define the
(normalized) ACF at lag $\tau$ for a signal $I(t)$ as \begin{equation}
  \mbox{ACF}(\tau) = {\int I(t) I(t+\tau) dt \over \int I^2(t) dt},
\label{acf} \end{equation} so that ACF$(0) = 1$.
The behavior of the ACF needs a closer look to understand what can
and cannot be inferred about the timescales of a signal.  For
illustration, consider the toy train of Gaussian spikes (Fig.~\ref{toy.acf})
which is characterized by two timescales; namely, the baseline width
$t_\mu$ of each spike, and the distance between two successive spikes,
\ie the period $P_\mu$ of the spike train.
The gap $g = P_\mu-t_\mu$ between two successive spikes is assumed to be non-negative.
When $t_\mu \le g$, it can be argued that the first ACF minimum occurs at lag $= t_\mu$, and
the first ACF maximum occurs at lag $=P_\mu$.  On the other hand, when
$g > t_\mu$, successive minima and maxima in the ACF are separated by
the period $P_\mu$ of the spike train.  In this regime, the ACF
carries no information about the width of a spike, and we expect that
the lag at the first maximum (\ie $P_\mu$) $= 2$ $\times$ the lag at
the first minimum (\ie $t_\mu$).  Conversely, any indication of the
relationship $P_\mu = 2 \times t_\mu$ in the inferred timescales may
therefore be interpreted as suggesting the lack of complete
information about the two timescales.

To motivate the methodology we are about to develop for inferring
microstructure timescales from individual subpulses in a pulsar data
set, consider four illustrative subpulses from PSR B0525+21 together
with their ACFs (Fig.~\ref{zoo}).  In the left column of this figure,
black curves represent probable candidates for the signal in the data
time series (gray scatter), red curves represent probable candidates for the
smooth trend (\ie \emph{envelope}) in the pulse, and blue curves
represent probable candidates for the microstructure (obtained after
removing the probable envelope from the probable signal).  In the
right column of this figure, the gray curves represent the ACF for the
whole-pulse time series, whereas the blue curves represent the ACF for the
microstructure feature.  Note that while the ACF for the whole-pulse
time series must be non-negative at all lags (because pulse time series are
by-and-large positive, barring noise), the ACF for the microstructure
feature may be negative (because the microstructure signal
fluctuates on both sides of 0).
In Fig.~\ref{zoo}, the time series for pulse 179 shows
strong microstructure oscillations superimposed on top of a smooth
envelope, which translate into weak but clear oscillations in the
descending part of the subpulse ACF close to lag 0.  For this pulse,
we see that the first minimum in the subpulse ACF (gray curve in the
ACF plot) agrees with the first minimum in the ACF for the
microstructure feature (blue curve in the ACF plot).  This pulse is
an example where the conventional method of identifying microstructure
timescales using the first minimum in the ACF will work flawlessly
because the first minimum in the subpulse ACF is numerically
well-defined.  The time series for pulse 37 again shows a highly
fluctuating signal characteristic of the microstructure.  However, the
first ACF minimum is barely visible in the subpulse ACF.  In
contrast, the first minimum is clearly identifiable in the
microstructure ACF for this pulse.  The time series for pulse 21 again
shows a clearly identifiable microstructure feature.  However, it
nevertheless leads to a featureless subpulse ACF which makes it
impossible for the conventional method to identify a microstructure
timescale from it.  The ACF for the microstructure feature, in contrast,
has a clearly identifiable first minimum.  Pulse 62 is an example
where, with the exception of one or two clear but weak fluctuations in 
the time series, there may not be any significant/detectable
microstructure feature.

These illustrative pulses suggest that the signal hidden in the data
can be modeled as consisting of an \emph{envelope} feature
characterized by relatively strong long-time smooth variations, and a
\emph{microstructure} feature characterized by relatively weak
short-time quasiperiodic variations that ride on top of the envelope.
Hence, with reference to the use of ACFs for identifying timescales, we
expect the following two confounding situations:
\begin{enumerate}
 \item
 The ACF of the complete pulse may be dominated by power at low
 frequencies, \ie by the envelope feature, possibly masking any
 short-time modulations (\eg pulses 37 and 36 in Fig.~\ref{zoo}).
 \item
 Noise in the data may result in additional random oscillations in the
 ACF, making it numerically difficult to identify genuine minima,
 slope changes, etc., in the ACF for the whole subpulse.
\end{enumerate}
This suggests that if the noise and the
envelope feature are extracted from the data to arrive at the
microstructure feature, then the microstructure feature can be used meaningfully to estimate
microstructure timescales from its ACF in the conventional manner.

Therefore, a solution to the problem of estimating microstructure
timescales lies in correctly estimating and removing the noise as well
as the long-time smooth modulation (\ie envelope) from a subpulse
time series.  Our methodology for estimating microstructure timescales
from individual subpulses thus consists of the following steps.
\begin{enumerate} 
\item The average pulse profiles for many pulsars
are often seen to be composed of multiple emitting components.  As was
pointed out in some earlier studies (\eg \citealt{1981SvA....25..442S}; 
\citealt{1983SvA....27..169S}; \citealt{2002A&A...396..171P}),
the individual components
can have different microstructure timescales.  Hence, we isolate
individual pulse components wherever possible before estimating
timescales.  Individual components are identified from the average
pulse profile in the form of time windows (shown as gray sections in 
the figures in Sec.\ \ref{a_avprof}) relative to the beginning of
the profile.  These windows are then applied to each single pulse in the
data set to separate it into its components.  For pulsars where
multiple components could not be identified, we used whole pulses.
Further, we restrict our analysis only to pulses with high S/N, where
S/N is estimated as the ratio of the peak signal to the off-pulse
noise.  In Table~\ref{tab1}, 24 pulsars marked in bold face are the
ones with sufficient single-pulse S/N and hence were suitable for
timescale analysis, while the others are not used. The pulsar
rotational parameters for these 24 pulsars are given in Table~\ref{tab1a}.

 \item 
RFI-induced outlier values in subpulse time series data may
influence timescale estimates.  We therefore identify potential
outliers in subpulse time series data using a heuristic method, and
replace them with appropriate values resembling their respective local
neighborhoods in the time series (Sec.\ \ref{a_outliers}).  The
decision to ``curate" outliers in this fashion is made on a case-to-case basis through
preliminary visual inspection of each pulsar data set.

 \item
 We de-noise every subpulse using a model-independent smoothing
 spline fit (Sec.\ \ref{a_fit}) that adapts to the shape of the subpulse optimally.  
 The fit thus obtained is taken to be a smooth noise-free version of the subpulse data
 (\eg the black curves in Fig.~\ref{zoo} left column).

 \item 
We estimate the envelope for each subpulse via kernel regression
(Sec.\ \ref{a_envelope}) using a heuristic smoothing timescale.
The difference between the fit and this envelope is the estimated microstructure.
Note that the smoothing timescale (referred to as smoothing bandwidth $h$ later
on) is the most critical parameter for our methodology.
Further details can be found in Sec.\ \ref{results}. 

 \item The nature and strength of the microstructure feature can
 vary considerably among subpulses.  Specifically, the microstructure 
 feature may be nearly absent or at best weak in some of the 
 subpulses.  We identify subpulses with weak microstructure using 
 the \emph{degrees of freedom} (DF) for the smoothing spline fit (Sec.~\ref{a_fit}),
 because lower DF values correspond to smoother fits and 
 weaker microstructures.
 For better estimation of microstructure timescales,
 we reject subpulses with DF in the lower 10-20 percentiles 
 of the DF distribution for that data set
(See Supplementary Table~\ref{tabC1} for the percentile cut-off for each data set). 

 \item For the selected pulses, the two characteristic microstructure
 timescales (width $t_\mu$ and period $P_\mu$) are estimated respectively
 as the first minimum and the first maximum in the ACF of the estimated
 microstructure for a subpulse.  Because of the substantial
 variation in microstructure characteristics across subpulses, the
 timescales thus obtained need to be further summarized or analyzed
 statistically.  Specifically, for each pulsar, we report the median
 timescale together with the median absolute deviation (MAD) as a
 measure of spread.  We also explore the nature of any
 relationship between $t_\mu$ and $P_\mu$.  
\end{enumerate} 
The microstructure timescale analysis pipeline developed in this paper is
implemented in the {R} \citep{R2014} open source statistical computing
environment. Detailed pulsarwise results is available as a tarball in:
\url{ftp://wm.ncra.tifr.res.in/dmitra/hires\_analysis\_pdf.tar.gz} 
or \url{http://cms.unipune.ac.in/reports/tr-20150223/}.

\section{Results}
\label{results}

Fig.~\ref{out_r1} and \ref{out_r2} depict typical results from our
analysis package.  Fig.~\ref{out_r1} shows how the technique is
applied to a subpulse (see caption for description).  We estimate the
microstructure periodicity $P_{\mu}$ as the first peak in the ACF of
the microstructure signal (blue line), and show the results for three
envelope smoothing bandwidths $h$ because the estimated timescales are
sensitive to the envelope bandwidth, where $h$ is the smoothing window
size in units of the total number of points in a subpulse (see
Sec.\ \ref{a_envelope} for details).  In the absence of any
theoretical prediction for microstructure timescales, it is difficult
to choose an appropriate value for $h$ from first principles.  Hence,
we used timescale estimates from earlier microstructure studies in the
literature to guide us, together with crucial properties of the
subpulse data to converge upon a range of reasonable $h$ values, as
follows.  For every pulsar, we identified a few highly
fluctuating/modulating pulses (like examples shown in Fig.~\ref{fig1})
where the $P_{\mu}$ estimated from first peak in the ACF is easily
discernible.  This gave us a range of $h$ values over which the
$P_{\mu}$ value remained unchanged.  Using a few representative $h$
values in this range we found the distributions of $P_{\mu}$.  The
median value of these $P_{\mu}$ distribution (or $t_{\mu}$ found from
first dip in the ACF) appeared to be in good agreement with available
estimates in the literature.  We found that for all pulsars, $h$
larger than 0.1 produced envelopes that were too smooth, whereas $h$
less than 0.05 led to envelopes that traced the subpulse data too
closely.  All our results are hence quoted for three representative
bandwidths, namely, 0.1, 0.075 and 0.05, as illustrated in
Fig.~\ref{out_r2}.  The median of the $P_{\mu}$ histogram is given in
columns 5, 6 and 7 of Table \ref{tab2} where the spread of the $P_\mu$
distribution is estimated using the median absolute deviation (MAD).
Column 2 gives the longitude range used to define the subpulses for
microstructure analysis, and these can be identified from the average
pulse profiles presented in Sec.\ \ref{a_avprof}.

The top panel of each plot in Fig.~\ref{out_r2} deserves attention. The plot depicts 
the relationship between the values of the first dip ($t_{\mu}$) along the $x$-axis and 
the first peak ($P_{\mu}$) along the $y$-axis as estimated from the ACFs of the 
microstructure. The three dotted lines correspond to 
$P_{\mu}=t_{\mu}, P_{\mu}=2t_{\mu}$ and $P_{\mu}=4t_{\mu}$
respectively. Most of the points in this figure (and also for other pulsars) are seen to cluster
around the $P_{\mu}=2t_{\mu}$ line, indicating that the microstructure is unresolved and
hence the width $t_{\mu}$ cannot be determined. 

Before we close this section, it is important to compare our analysis method with 
earlier studies. Majority of the studies use the first dip or peak of the ACF
to estimate microstructure timescales (\eg \citealt{1981SvA....25..442S}; 
\citealt{1972ApJ...177L..11H}). However, as mentioned earlier, 
the broad subpulse power as well as noise in the subpulse make it difficult
to extract the timescales numerically.  Our method is designed to alleviate
these problems by first de-noising the subpulse profile and then removing a smooth
envelope from it. Other studies have used the power spectrum to estimate 
timescales corresponding to quasiperiodic microstructures 
(\eg \citealt{2002A&A...396..171P}; \citealt{1998A&A...332..111L}). The slow-varying envelope in these 
cases appears as low-frequency power in the spectrum. For example,
\citet{2002A&A...396..171P} constructs the ACF by cross-correlating the voltages
from two different spectral channels, while \citet{1998A&A...332..111L} 
use the derivative of the power spectrum, although noise in single pulses
may still affect their analyses.    
The strength of our method lies in its ability to recover
low-amplitude intensity modulations in a subpulse, isolating the microstructure from the subpulse time series.
Consequently, we find that majority of the subpulses have microstructure emission superposed on a smooth broad emission (envelope).

\section{Distribution of timescales for all Stokes parameters and the $P_{\mu}$-$P$ relation}
\label{sec4}
\citet{1979AuJPh..32....9C}, based on 10 pulsars observed at 430 MHz
in the pulsar period range of 0.2 to 3.7 second, found the
relationship t$_{\mu} \sim 10^{-3} P$ between the microstructure
timescale $t_{\mu}$ and the pulsar period $P$.
\citet{2002MNRAS.334..523K} reestablished this relationship as
t$_{\mu} (\mu s) \sim(600\pm100) P (s)^{1.1\pm0.2}$ using 12 pulsars
where $t_{\mu}$ estimates were available in the literature for
different frequencies.  With our current set of observations we are in
a position to revisit this relationship using 24 pulsars at L Band.
Since broad microstructures cannot be resolved, we believe that the
periodicity $P_{\mu}$ is a more relevant quantity instead of
$t_{\mu}$.  Fig.~\ref{per_pmu} plots the the median value of $P_{\mu}$
against period $P$.  The top panel corresponds to the total intensity,
the middle one to the linear polarization, and the bottom one to the
circular.  The $P_{\mu}$ points are estimated from the median value of
the histogram of the first peak in the ACF, and the error bars reflect
the median absolute deviation of the $P_{\mu}$ distribution.  A linear
fit of the form $P_{\mu} \mbox{ (msec)} = m P \mbox{ (msec)} + c $ was
performed using the median $P_{\mu}$ values for stokes I corresponding
to $h = 0.1, 0.075$ and 0.05 separately, which yielded $(m,c)$ values
as $(1.3\times 10^{-3}, 0.04), (1.1\times 10^{-3}, 0.08)$ and
$(0.9\times 10^{-3}, 0.098)$ respectively. These fits are shown in
Fig.~\ref{per_pmu}.  We note that the fit is merely demonstrating the
tendency that median $P_{\mu}$ increases with increasing pulsar period
$P$, although a large spread of $P_{\mu}$'s exists for all pulsars.

Correlation of $P_{\mu}$ with other pulsar rotational parameters like
period derivative, characteristic age and surface magnetic field did not 
show any significant trends. Although a weak anticorrelation was seen (not shown here) between 
the slowdown energy ($\dot{E}$) and $P_{\mu}$, more data is needed to 
establish this. 

\section{How does broad microstructure average to form a subpulse?}
\label{sec5}

It was evident from early polarization observations (\eg \citealt{1969Natur.221..724C};
\citealt{1975ApJ...196...83M}) that single pulses are highly linearly polarized,
and that the circular polarization changes sign or handedness within a pulse.
Based on high-quality single-pulse polarimetry, \citet[hereafter MGM09]{2009ApJ...696L.141M} have
showcased a collection of subpulses from an assorted set of pulsars, where the subpulses had a Gaussian 
shape, with close to 100\% linear polarization and associated with sign-changing circular.  
One such example is shown in the left panel of Fig.~\ref{fig3}.

For a long time this emission pattern kindled great interest in the pulsar community (see 
\eg \citealt{1987ApJ...322..822M}; \citealt{1990ApJ...352..258R}), because this remarkably resembles the 
emission pattern due to single-particle vacuum curvature radiation as the observer cuts 
across the emission cone (\ie the 1/$\gamma$ cone).  The mathematical formulation for 
obtaining Stokes parameters due to curvature radiation in vacuum for a single relativistically 
moving particle has been explicitly calculated by several authors (\eg \citealt{1990A&A...234..269G}; 
\citealt{1990A&A...234..237G}; \citealt{2002ApJ...566..365A}). 
Let us recall the polarization properties of single-particle 
curvature radiation in vacuum based on the description given by \citet{1990A&A...234..237G}, \citet{1990A&A...234..269G}.  
Consider a charged particle moving relativistically with Lorentz factor $\gamma$, along
super-strong magnetic field lines, with a radius of curvature $\rho$.  The particle will emit 
curvature radiation beamed in the forward direction.  The Fourier components of the electric 
fields can be decomposed into parallel ($\varepsilon_{\parallel}$) and perpendicular 
($\varepsilon_{\perp}$) components to the plane of the curved magnetic field lines 
(see eqs. 7 and 8 of \citealt{1990A&A...234..237G}).  When the radiation is viewed as a cut through 
the emission beam, the $\varepsilon_{\perp}$ has a sinusoidal form, $\varepsilon_{\parallel}$ 
is positive and has a relative phase-shift of 90$\degr$ forming an orthogonal basis.  
The radiation polarized in the parallel and perpendicular direction are also known as the 
ordinary (O) and extraordinary (X) modes.  The power in curvature radiation is dependent 
on the frequency of emission, and achieves its maximum at $\nu_{c}$, the characteristic 
frequency which is given by $\nu_{c} = 1.25 \gamma^{3} \nu_{0}$ in the 
vacuum case (See fig. 2 of \citealt{1990A&A...234..269G}).  The ratio of the two components as well 
as the width of the pulse is dependent both on the viewing angle $\varphi$ and the frequency 
of emission (see figs. 3, 4 and 5 of \citealt{1990A&A...234..269G}).  The width of the pulse is
approximately equal to $\delta = 1/\gamma$ at $\nu_{c}$.  The power in the O mode is 
seven times stronger than the X mode in vacuum (\citealt{1975clel.book.....J}; 
\citealt{2004ApJ...600..872G}).

Our aim here is to look for signatures of the aforementioned vacuum curvature radiation 
in the data.  First, let us note that the subpulse width cannot be associated with the elementary 
1/$\gamma$ emission cone, because when observed with higher time resolution, the subpulse 
splits into microstructures.  In an attempt to reproduce this subpulse radiation pattern,
\citet{1990A&A...234..237G} and \citet{1993A&A...272..207G} used the shot-noise model of pulsar emission, 
where the subpulse is an incoherent addition of smaller bunches, with each bunch emitting 
like a single-particle vacuum curvature radiation.  The subpulse shape results from the 
intensity gradient due to the charge density variation of the bunches across the subpulse.  
These authors demonstrated that in the presence of intensity gradients, the resulting emission 
will have a net circular polarization as observed in subpulses.  The beams from the bunches 
do not overlap, so the linear polarization remains high.  Assuming $\gamma \sim 100$, and therefore 
an angular size $1/\gamma\sim 0.6^{\circ}$ for the emitting bunches, they simulated several 
pulses that appear similar to the subpulses showcased by MGM09.  These results were 
also recently confirmed by \citet{2010ApJ...710...29G}.  A prediction of this model is that if a subpulse 
can be resolved into microstructures (\eg see fig. 3 of \citealt{1993A&A...272..207G}), then the microstructure 
should have features which mimic the single-particle vacuum curvature-radiation pattern.  To test 
this theory we need examples of subpulses where the microstructures are highly resolved.  
Note that the subpulses chosen should be highly linearly polarized such that they do not 
suffer from any depolarization effects due to orthogonal polarization moding\footnote{It is 
  important to touch upon a subtle point regarding sign-changing circular
  polarization in pulsar subpulses.  In some cases, at the zero-crossing
  point of the sign-changing circular, the linear polarization
  drops significantly and the polarization position angle (PPA) gets associated with an orthogonal
  jump.  However, for testing the vacuum curvature-radiation signatures,
  such subpulses are undesirable as they do not comply with the basic
  property of vacuum curvature radiation where, at the zero-crossing
  point of the circular, the fractional linear polarization has to be close to
  100\%.} or due to incoherent addition from overlapping
$1/\gamma$ emission beam (see MGM09).

\begin{figure*}
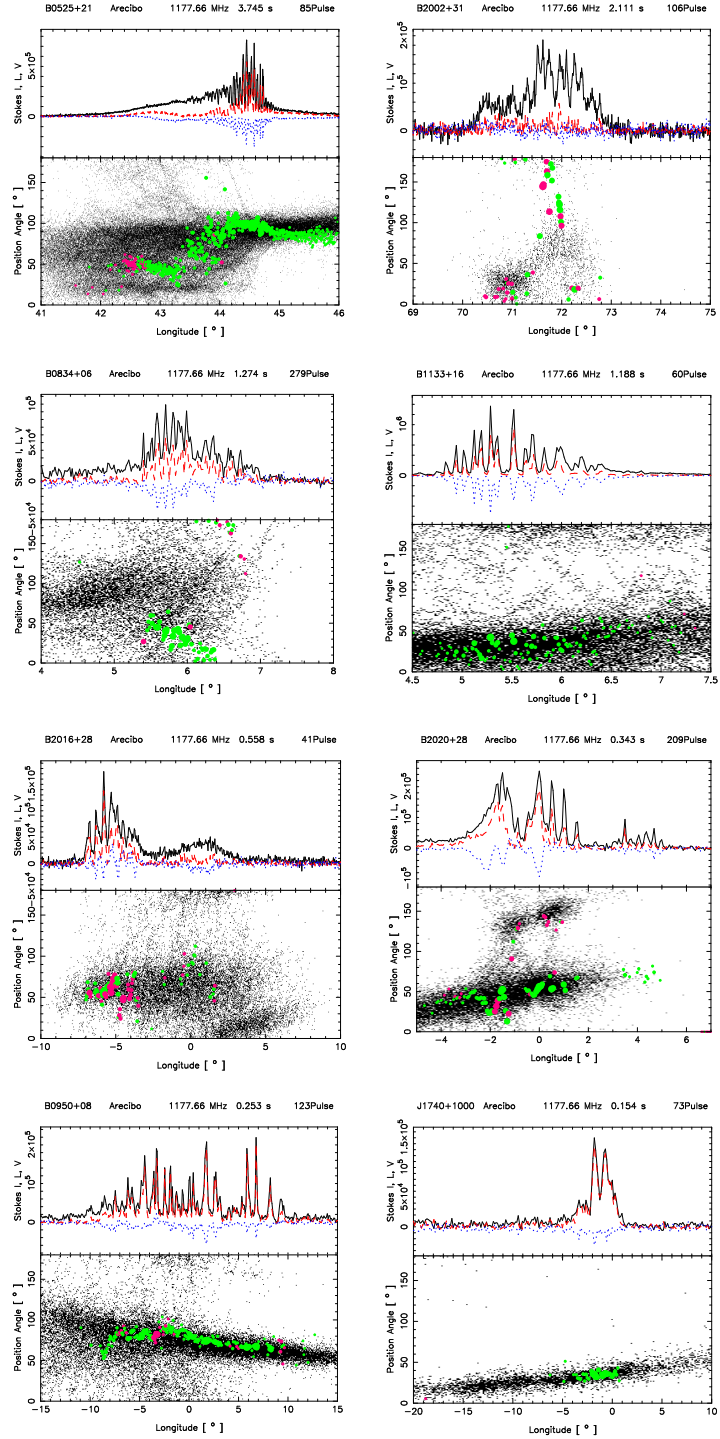

\begin{center}
\begin{tabular}{@{}lr@{}}
 {\mbox{\includegraphics[width=45mm,angle=-90.]{B0525_p85.ps}}}& 
 {\mbox{\includegraphics[width=45mm,angle=-90.]{B2002_p106.ps}}}\\ 
 {\mbox{\includegraphics[width=45mm,angle=-90.]{B0834_p279.ps}}} & 
 {\mbox{\includegraphics[width=45mm,angle=-90.]{B1133_p61.ps}}} \\
 {\mbox{\includegraphics[width=45mm,angle=-90.]{B2016_p41.ps}}}&
 {\mbox{\includegraphics[width=45mm,angle=-90.]{B2020_p209.ps}}}\\
 {\mbox{\includegraphics[width=45mm,angle=-90.]{B0950_p123.ps}}}&
 {\mbox{\includegraphics[width=45mm,angle=-90.]{J1700_p73.ps}}}\\
\end{tabular}
\caption{A collection of subpulses are shown in full polarization for various of the pulsars 
observed with a time resolution of 59.5 $\mu$sec.  In each plot the top panel shows Stokes 
$I$, $L$ and $V$ in black, red and blue lines respectively.  The polarization position angle (PPA) histogram of the entire 
pulse sequence is shown as a ``dotty plot'' display of qualifying samples in the bottom panel.  
The average PPA of the subpulse is plotted in green when the circular polarization is negative 
and magenta when positive.  The ringing/modulating structures seen in the subpulses are 
microstructures.  Note that a certain periodicity (or quasiperiodicity) is apparent in the 
microstructures, and all the Stokes parameters fluctuate in the same manner with the same 
width and periodicity.}
\label{fig1}
\end{center}
\end{figure*}

\begin{figure*}
\begin{center}
\begin{tabular}{@{}lr@{}}
 {\mbox{\includegraphics[width=75mm,angle=-90.]{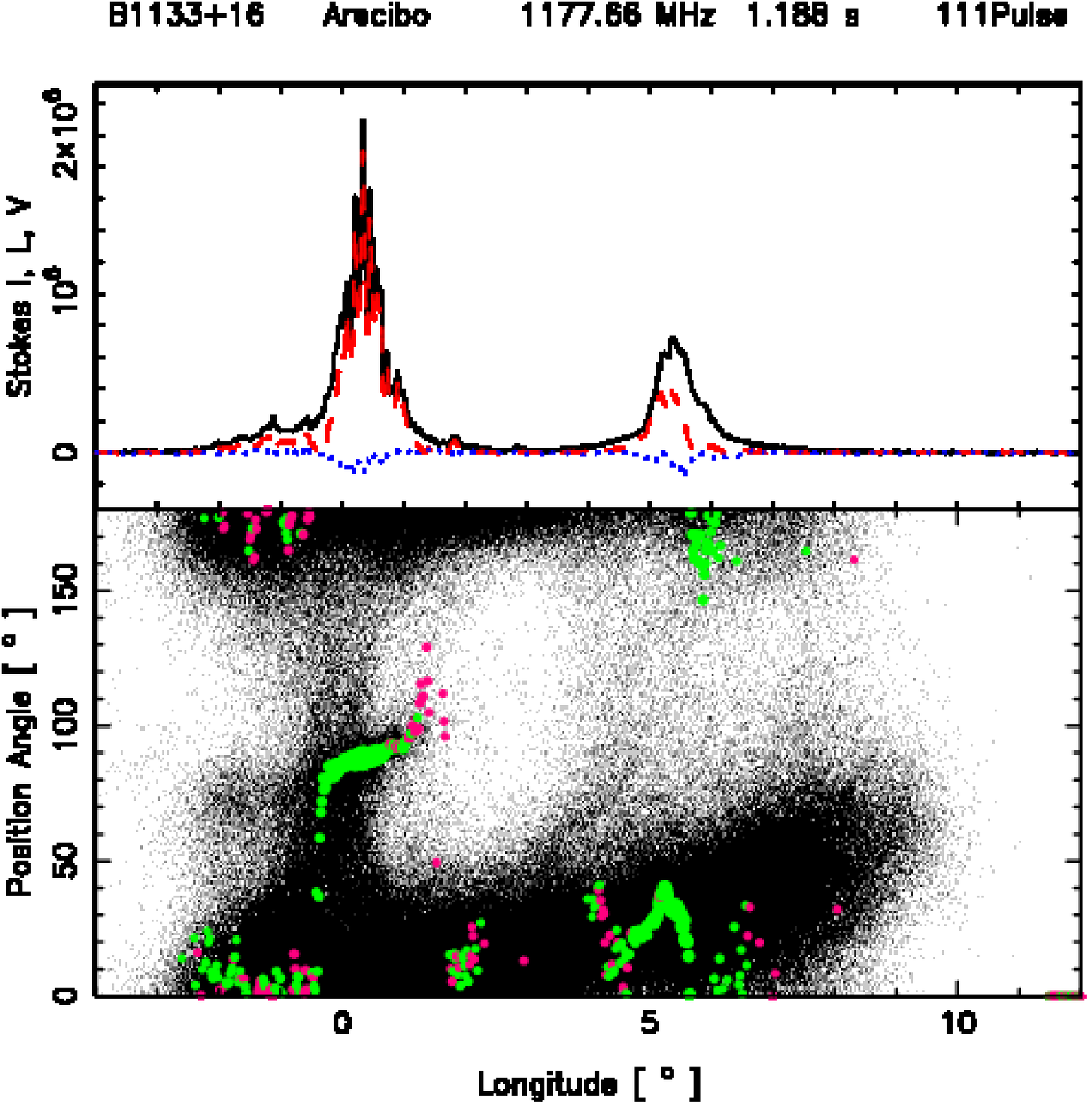}}}& 
 {\mbox{\includegraphics[width=75mm,angle=-90.]{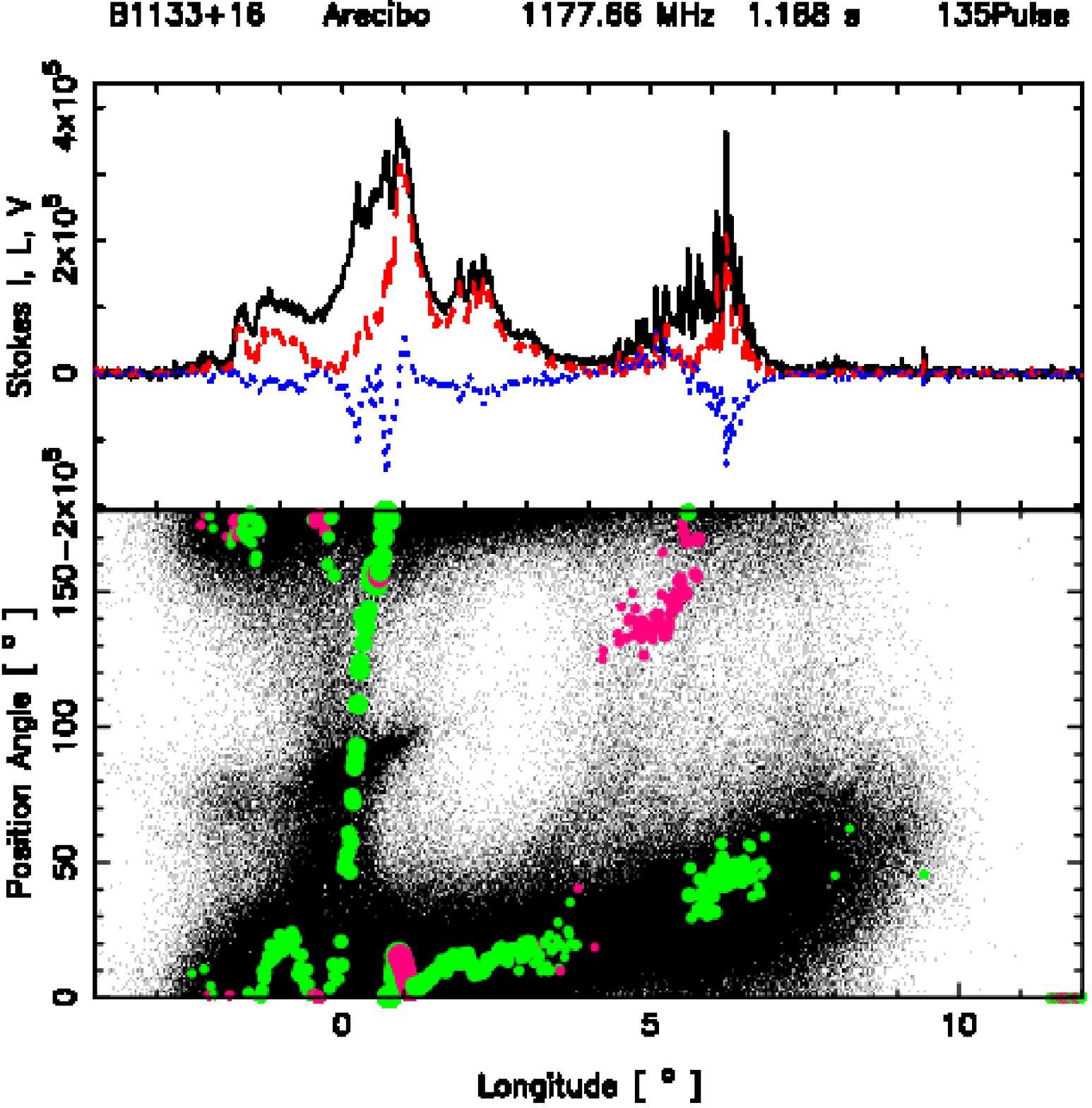}}}\\
\end{tabular}
\caption{Two single pulses of PSR B1133+16 (plot description same as given 
in the caption of Fig.~\ref{fig1}), illustrating 
that microstructures occur within both polarization modes.  In the left hand display 
around 0$\degr$ longitude the pulse component showing the microstructure 
feature corresponds to the weaker secondary polarization mode; whereas in 
the right hand one at about 6$\degr$ the microstructure corresponds to the 
primary polarization mode.}
\label{fig2}
\end{center}
\end{figure*}

\begin{figure}
 \includegraphics[height=\textwidth,angle=-90]{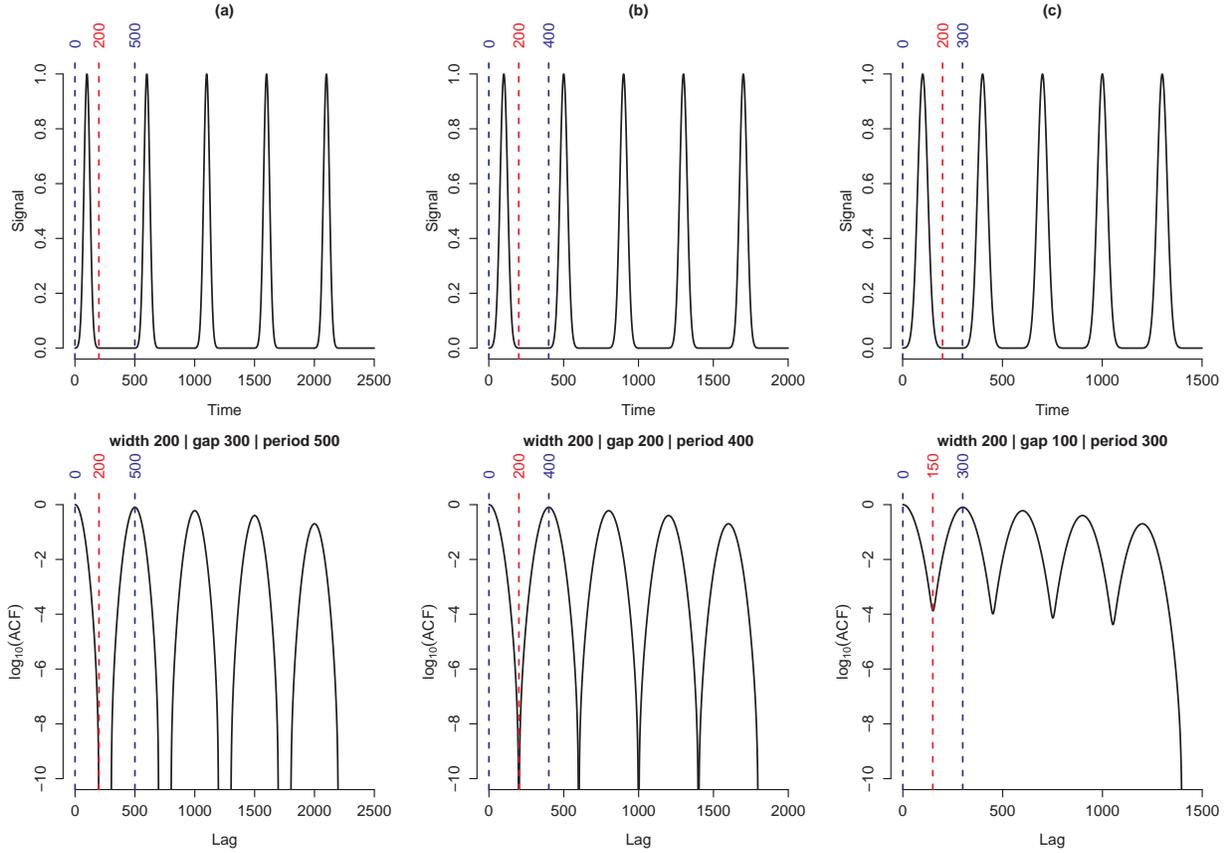}
 \caption{\label{toy.acf} A toy example illustrating the relationship
 between the width of a spike, period of a spike train, and the first
 minimum and maximum in the ACF. In the top row, each spike is a
 truncated Gaussian with baseline width = 200, and the spike train is
 a sequence of equidistant spikes separated with the same gap. The period
 of the spike train therefore equals the widths of a spike + gap between
 successive spikes. (a) width $<$ gap; (b) width $=$ gap; (c) width
 $>$ gap. In all cases, the first ACF maximum is at lag = period. For
 (a) and (b), the first ACF minimum is at lag $=$ width, and the ACF
 can be used to infer both the width and period of the spike train. For
 (c), the first ACF minimum is at lag $=$ period$/2$, and the ACF
 carries no information about the width. The spike train is a
 caricature of the microstructure feature of a subpulse, and
 illustrates what can be inferred about timescales using the ACF. Note
 that the bottom row plots log$_{10}$(ACF) instead of the ACF so as to
 accentuate the ACF minima.}  \end{figure}

In the left hand panel of Fig.~\ref{fig3}, a subpulse of PSR B2020+28 is shown which 
is highly linearly polarized and is associated with a sign-changing circular polarization, 
similar to the pulses showcased by MGM09.  The right hand panel shows the high time 
resolution data for the same subpulse, and clearly resolved microstructures are seen.  
As expected, the linear polarization of the individual microstructures are close to 100\% 
in order to produce the highly polarized subpulse.  The circular polarization below the 
microstructure feature does not show any sign reversals. Rather it has the same 
width (and also periodicity) as the total intensity and linear polarization.  In fact, this 
description is true for all subpulses where sufficiently resolved microstructures are 
seen.  Fig.~\ref{fig1} shows several such examples, and very clear examples where 
Stokes $V$ has the same $P_{\mu}$ as Stokes $I $ and $L$ are shown in Fig.~\ref{fig4}.  
This feature is also borne out from the statistical analysis of the data set where the 
median $P_{\mu}$ values of Stokes $I$, $L$ and $V$ are similar as can be seen in 
Table~\ref{tab2}.  Thus, based on these observations, one is compelled to conclude that 
radiation pattern of resolved microstructures are not consistent with single-particle
curvature radiation of charged bunches in vacuum.  In the next section, we will discuss 
how these findings corroborate our current understanding of the pulsar emission 
mechanism.

\section{Pulsar microstructure and its implications for the coherent radio emission mechanism}

Coherent pulsar radio emission can be generated by means of either a
maser or coherent curvature mechanism (\citealt{1975ARA&A..13..511G};
\citealt{1991MNRAS.253..377K}; \citealt[hereafter
  MGP00]{2000ApJ...544.1081M}).  There is general agreement that this
radiation is emitted in curved magnetic field-line planes due to
growth of plasma instabilities in the strongly magnetized
electron-positron plasma well inside the light cylinder.  Many
observational constraints on the emission altitude suggest that the
emitted radiation detaches from the ambient plasma at altitudes of
about 500 km above the stellar surface which is typically less than
10\% of the light cylinder radius $R_{LC}=Pc/2\pi$ (\eg
\citealt{1978ApJ...222.1006C}; \citealt{1991ApJ...370..643B};
\citealt{1993ApJ...405..285R}; \citealt{1997MNRAS.288..631K};
\citealt{1999A&A...346..906M}; \citealt{2001ApJ...555...31G};
\citealt{2004A&A...421..215M}; \citealt{2009MNRAS.393.1617K}).  The
magnetic field at these altitudes is so strong that the plasma is
constrained to move only along the magnetic field lines.  Under such
conditions, the two-stream instability is the only known instability
that can develop, and the growth of this plasma instability leads to
the formation of charged relativistic solitons, which can excite
coherent curvature radiation in the plasma (\eg
\citealt{1998MNRAS.301...59A}; \citealt{2000ApJ...544.1081M};
\citealt{2004ApJ...600..872G} GLM04 hereafter).

The basic requirement of the theory of coherent radio emission is the presence of a non-stationary 
inner acceleration region.  \citet{1975ApJ...196...51R} (hereafter RS75) were amongst 
the first to explore the conditions in the pulsar magnetosphere that lead to the formation 
of an inner accelerating region (also known as an inner vacuum gap, IVG) just above 
the pulsar polar cap where the free flow of charge is hindered, thus resulting in the 
formation of an inner vacuum gap with a large potential difference.  The physics of inner 
vacuum gap formation can be found in RS75, and here we point out that the original 
RS75 inner vacuum gap was not consistent with x-ray observations and radio subpulse 
drifting observations, and currently it is thought that a partially screened inner vacuum gap 
exists close to the neutron star surface \citep{2003A&A...407..315G}.  The process 
that generates non-stationary flow of charges in the gap is as follows.  The IVG is initially 
discharged by a photon-induced electron-positron pair-creation process in the strong 
curved magnetic field.  The height ($h_g$) of the gap stabilizes typically at about a few 
mean-free paths of the pair-production process.  The electric field in the gap causes 
separation of the electrons and positrons (also known as primary particles), where the 
electrons bombard onto the neutron-star surface, and the positrons are accelerated away 
toward the outer magnetosphere.  As these primary particles move they accelerate and create 
further secondary particles resulting in a pair-creation cascade.  The outflowing particles 
are hence composed of primary particles with Lorentz factors of about 10$^6$ and secondary 
plasma clouds with a distribution of Lorentz factors from about 10 to 1000, but peaking 
around 300--500.  This process eventually occurs over a set of magnetic field lines, and 
this overall spread of the discharging process is called a spark.  As the electron moves 
towards the polar cap it experiences $E \times B$ drift, causing the sparking discharge 
to move slowly across the polar cap.  RS75 envisaged that the inner vacuum gap is 
discharged in the form closely packed isolated sparks, where the lateral size and distance
between the sparks is equal to the gap height.  A plasma column is associated with 
each spark, and in the radio-emission region coherent spark-associated plasma 
columns radiate and generate the observed subpulses in pulsars.  The sparking 
process continues for a time $t_s$ until the whole potential in the gap is screened out, 
after which the plasma empties the gap in a time $t_g = h_g/c$ and once again the 
sparking process starts almost in the same place on a timescale $\sim t_g << t_s$ 
\citep{2000ApJ...541..351G}.  This process results in a non-stationary flow of plasma clouds 
associated with every spark anchored to a specific location on the polar cap.  The physics 
of the sparking discharge process is still a topic of active research, however most studies 
suggest that in normal pulsars $h_g$ is of the order of about 50--100 m, and to screen the 
gap potential around 50 pair creations are needed; and hence $t_s \sim 50 h_g/c \sim 10\mu$ sec.  
Note that during the time $t_s$ the $E \times B$ drift causes the spark to move only very 
slightly across the polar cap.  Two successive plasma clouds, with each cloud extending 
for about 50$h_g$ and separated by a distance $h_g$, eventually overlap when the fast 
moving particles of the second cloud catch up with the slow moving particles of the first 
cloud, leading to the development of the two-stream instability.  The instability in the plasma 
generates Langmuir plasma waves and the modulational instability of the Langmuir waves 
leads to the formation of a large number of charged solitons which can excite extraordinary
(X) and ordinary (O) modes of curvature radiation in the plasma cloud as shown by MGP00 
and GLM04.  The condition for coherent emission requires the wavelength of emission 
to be greater than the intrinsic soliton size, \ie $\omega < 2\sqrt{\gamma_s} \omega_p$ 
where $\omega_p = (4\pi e^2 n/m_e)^2$ is the plasma frequency (see MGP00, GLM04), 
and $\gamma_s$ is the Lorentz factor of the secondary plasma.  The emerging radiation 
is an incoherent addition of the radiation emitted by each soliton.

We now discuss aspects of the soliton coherent-curvature radiation theory that relate 
to the two important microstructure behaviors discussed in this paper, namely the periodic 
structures observed in pulsar subpulses, their linear dependence with pulsar period and 
the absence of signatures of vacuum curvature radiation in pulsar microstructures.  
Obviously the timescales involved in the radiation process are far too small to be seen in 
our 59.5 $\mu$sec observations, hence we rule out that the periodic microstructures 
correspond to emission from individual plasma clouds or solitons\footnote{Note that
  in this context the findings of Popov \etal\ (2002) are rather
  interesting as based on their 62.5 nanosec observations they report
  that the shortest microstructure widths are about 2$\mu$s to 10$\mu$s, which
corresponds to the time over which the sparking discharge
takes place.}. Another way of sampling a periodic structure would be if 
temporal modulation of the emission exists in the emission
region.  There are theoretical studies in the literature which attempt to interpret 
microstructure based on nonlinear plasma effects leading to temporal modulation
of the pulsar emission (\eg \citealt{1983Ap&SS..97....9C}; \citealt{1998ApJ...506..341W}; \citealt{1993MNRAS.264..940A}); 
however these models are mostly developed for emission where $\omega \geq 2\sqrt{\gamma_s} \omega_p$,
whereas coherent emission requires $\omega < \omega_p$.

The other important aspect of our observations is the absence of a signature of vacuum 
curvature radiation in pulsar microstructures.  
The coherent curvature radiation by charged bunches called solitons excites the 
extraordinary and ordinary mode in the electron-positron plasma.  Due to the strong 
ambient magnetic field the X-mode does not interact with the plasma and hence can 
escape the plasma and reach the observer (MGM09, \citealt{2014ApJ...794..105M}) .  
The fate of the O-mode is however
unclear.  It has been suggested that it can be damped and gets ducted along the magnetic 
field lines, or if there are sharp plasma boundaries that it can escape as electromagnetic 
waves and reach the observer.  In any case, due to propagation effects in the plasma, the 
X and O modes separate in phase and hence they do not maintain any phase relation
as they emerge from the plasma. As a result the circular polarization can no longer 
have any sign-changing signature as in the vacuum case.  Also incoherent addition 
of a large number of solitons would tend to cancel any circular polarization, particularly 
in our observations.  Thus the origin of circular polarization still remains to be understood 
in this model of pulsar emission.

\section{Summary}
\label{discuss}
In this work we have found several distinguishing features of 
pulsar microstructures. These can be summarized as follows 
\begin{enumerate}
 \item Almost all subpulses show a quasiperiodic structure riding on 
top of a broad pulse envelope. In general this structure is weak, except 
in occasional cases when deep modulations are seen. 
However, the absence of microstructure in pulsar average profiles is a
clear indication that at any given longitude pulsar emission is
random and stochastic.

\item Microstructures are also seen to be associated with
both the orthogonal polarization modes. 

\item We did not find any case where microstructure widths are resolved. 
In other words, we find that $P_\mu \sim 2t_\mu$ (where the estimated width is $t_\mu$ 
found as the first dip in the ACF) is related to the periodicity $P_\mu$ (found as the 
first peak in the ACF) (see Sec.~\ref{sec2} for details).

\item For every pulsar the $P_\mu$ values extracted from a collection of subpulses 
have a distribution, and the median value of $P_\mu$ linearly increases with pulsar period (see
Sec.~\ref{sec4}).

\item All the Stokes parameters have the same periodicity---\ie they 
have same $P_\mu$ values (see Sec~\ref{sec5}). 
\end{enumerate}

A variety of analysis techniques to extract microstructure timescales
have been deployed in the past. The methods use either the ACF first dip and
first peak or use the peaks in the power spectra for estimating timescales.
However, these methods cannot effectively find the low-amplitude wiggles 
that are observed in subpulses.  We address this problem by de-noising the 
subpulse signal and subsequently removing a broad envelope. The ACF of 
the resultant signal is then used to estimate microstructure timescales. 

These quasiperiodic structures, which show a linear dependence on rotation
period, cannot be understood in terms of the current theories of pulsar 
emission. We also find that the linear and circular polarization follow the same 
distribution of timescales as the total intensity.  This, as we have argued 
strongly, rules out the possibility that microstructures are basic units of curvature 
radiation in vacuum.  We conclude that the microstructures within the pulsar signal 
reflect temporal modulations of plasma processes of the emission engine.
While this paper is an exercise in understanding the properties of pulsar 
microstructures, theoretical progress needs to be made to understand this 
unique phenomenon. 

\section{Acknowledgements}

\acknowledgments
We thank the anonymous referee for his/her thorough review and highly
appreciate the comments and suggestions which significantly improved
the paper.
We thank George Melikidze and Janusz Gil for useful discussions.  
We also thank M.\ Popov for his comments on the paper. 
DM would 
like to thank Cornell University for extending a visiting staff position at the 
Arecibo Observatory, during which many of the observations reported in this 
paper were carried out. We would like to thank Arun Venkataraman for his 
many helps and services as well as the other staff of Arecibo Observatory for 
support.  Much of the work was made possible by support from the US National 
Science Foundation grants 08-07691 and 09-68296.  Arecibo Observatory is 
operated by SRI International under a cooperative agreement with the National 
Science Foundation, and in alliance with Ana G.\ M\'endez-Universidad 
Metropolitana and the Universities Space Research Association.  This work 
made use of the NASA ADS astronomical data system.
MA was supported by the National Centre for Radio Astrophysics, TIFR, as a visiting scientist during 2011-13.

\bibliographystyle{apj}
\bibliography{micro}

\begin{thebibliography}{}
\expandafter\ifx\csname natexlab\endcsname\relax\def\natexlab#1{#1}\fi

\bibitem[{{Ahmadi} \& {Gangadhara}(2002)}]{2002ApJ...566..365A}
{Ahmadi}, P., \& {Gangadhara}, R.~T. 2002, \apj, 566, 365

\bibitem[{{Asseo}(1993)}]{1993MNRAS.264..940A}
{Asseo}, E. 1993, \mnras, 264, 940

\bibitem[{{Asseo} \& {Melikidze}(1998)}]{1998MNRAS.301...59A}
{Asseo}, E., \& {Melikidze}, G.~I. 1998, \mnras, 301, 59

\bibitem[{{Bartel} \& {Hankins}(1982)}]{1982ApJ...254L..35B}
{Bartel}, N., \& {Hankins}, T.~H. 1982, \apjl, 254, L35

\bibitem[{{Blaskiewicz} {et~al.}(1991){Blaskiewicz}, {Cordes}, \&
  {Wasserman}}]{1991ApJ...370..643B}
{Blaskiewicz}, M., {Cordes}, J.~M., \& {Wasserman}, I. 1991, \apj, 370, 643

\bibitem[{{Boriakoff}(1976)}]{1976ApJ...208L..43B}
{Boriakoff}, V. 1976, \apjl, 208, L43

\bibitem[{{Chian} \& {Kennel}(1983)}]{1983Ap&SS..97....9C}
{Chian}, A.~C.-L., \& {Kennel}, C.~F. 1983, \apss, 97, 9

\bibitem[{{Clark} \& {Smith}(1969)}]{1969Natur.221..724C}
{Clark}, R.~R., \& {Smith}, F.~G. 1969, \nat, 221, 724

\bibitem[{{Cordes}(1978)}]{1978ApJ...222.1006C}
{Cordes}, J.~M. 1978, \apj, 222, 1006

\bibitem[{{Cordes}(1979)}]{1979AuJPh..32....9C}
---. 1979, Australian Journal of Physics, 32, 9

\bibitem[{{Cordes} {et~al.}(1990){Cordes}, {Weisberg}, \&
  {Hankins}}]{1990AJ....100.1882C}
{Cordes}, J.~M., {Weisberg}, J.~M., \& {Hankins}, T.~H. 1990, \aj, 100, 1882

\bibitem[{{Craft} {et~al.}(1968){Craft}, {Comella}, \&
  {Drake}}]{1968Natur.218.1122C}
{Craft}, H.~D., {Comella}, J.~M., \& {Drake}, F.~D. 1968, \nat, 218, 1122

\bibitem[{Efromovich(1999)}]{Efromovich1999}
Efromovich, S. 1999, Nonparametric Curve Estimation: Methods, Theory, and
  Applications (New York, USA: Springer)

\bibitem[{{Everett} \& {Weisberg}(2001)}]{2001ApJ...553..341E}
{Everett}, J.~E., \& {Weisberg}, J.~M. 2001, \apj, 553, 341

\bibitem[{{Ferguson} \& {Seiradakis}(1978)}]{1978A&A....64...27F}
{Ferguson}, D.~C., \& {Seiradakis}, J.~H. 1978, \aap, 64, 27

\bibitem[{Frigge {et~al.}(1989)Frigge, Hoaglin, \& Iglewicz}]{FHI1989}
Frigge, M., Hoaglin, D.~C., \& Iglewicz, B. 1989, The American Statistician,
  43, pp. 50

\bibitem[{{Gangadhara}(2010)}]{2010ApJ...710...29G}
{Gangadhara}, R.~T. 2010, \apj, 710, 29

\bibitem[{{Gangadhara} \& {Gupta}(2001)}]{2001ApJ...555...31G}
{Gangadhara}, R.~T., \& {Gupta}, Y. 2001, \apj, 555, 31

\bibitem[{{Gil} {et~al.}(1993){Gil}, {Kijak}, \& {Zycki}}]{1993A&A...272..207G}
{Gil}, J., {Kijak}, J., \& {Zycki}, P. 1993, \aap, 272, 207

\bibitem[{{Gil} {et~al.}(2004){Gil}, {Lyubarsky}, \&
  {Melikidze}}]{2004ApJ...600..872G}
{Gil}, J., {Lyubarsky}, Y., \& {Melikidze}, G.~I. 2004, \apj, 600, 872

\bibitem[{{Gil} {et~al.}(2003){Gil}, {Melikidze}, \&
  {Geppert}}]{2003A&A...407..315G}
{Gil}, J., {Melikidze}, G.~I., \& {Geppert}, U. 2003, \aap, 407, 315

\bibitem[{{Gil} \& {Sendyk}(2000)}]{2000ApJ...541..351G}
{Gil}, J.~A., \& {Sendyk}, M. 2000, \apj, 541, 351

\bibitem[{{Gil} \& {Snakowski}(1990{\natexlab{a}})}]{1990A&A...234..237G}
{Gil}, J.~A., \& {Snakowski}, J.~K. 1990{\natexlab{a}}, \aap, 234, 237

\bibitem[{{Gil} \& {Snakowski}(1990{\natexlab{b}})}]{1990A&A...234..269G}
---. 1990{\natexlab{b}}, \aap, 234, 269

\bibitem[{{Ginzburg} \& {Zhelezniakov}(1975)}]{1975ARA&A..13..511G}
{Ginzburg}, V.~L., \& {Zhelezniakov}, V.~V. 1975, \araa, 13, 511

\bibitem[{Green \& Silverman(1994)}]{GS1994}
Green, P., \& Silverman, B. 1994, Nonparametric Regression and Generalized
  Linear Models: A Roughness Penalty Approach (New York, USA: Chapman and Hall)

\bibitem[{{Hankins}(1971)}]{1971ApJ...169..487H}
{Hankins}, T.~H. 1971, \apj, 169, 487

\bibitem[{{Hankins}(1972)}]{1972ApJ...177L..11H}
---. 1972, \apjl, 177, L11

\bibitem[{{Hankins} {et~al.}(2003){Hankins}, {Kern}, {Weatherall}, \&
  {Eilek}}]{2003Natur.422..141H}
{Hankins}, T.~H., {Kern}, J.~S., {Weatherall}, J.~C., \& {Eilek}, J.~A. 2003,
  \nat, 422, 141

\bibitem[{H\"ardle(1992)}]{Haerdle1992}
H\"ardle, W. 1992, Applied Nonparametric Regression (Cambridge, UK: Cambridge
  University Press)

\bibitem[{{Jackson}(1975)}]{1975clel.book.....J}
{Jackson}, J.~D. 1975, {Classical electrodynamics} (New York, USA: Wiley)

\bibitem[{{Jessner} {et~al.}(2010){Jessner}, {Popov}, {Kondratiev}, {Kovalev},
  {Graham}, {Zensus}, {Soglasnov}, {Bilous}, \&
  {Moshkina}}]{2010A&A...524A..60J}
{Jessner}, A., {Popov}, M.~V., {Kondratiev}, V.~I., {et~al.} 2010, \aap, 524,
  A60

\bibitem[{{Johnston} {et~al.}(2005){Johnston}, {Hobbs}, {Vigeland}, {Kramer},
  {Weisberg}, \& {Lyne}}]{2005MNRAS.364.1397J}
{Johnston}, S., {Hobbs}, G., {Vigeland}, S., {et~al.} 2005, \mnras, 364, 1397

\bibitem[{{Kazbegi} {et~al.}(1991){Kazbegi}, {Machabeli}, \&
  {Melikidze}}]{1991MNRAS.253..377K}
{Kazbegi}, A.~Z., {Machabeli}, G.~Z., \& {Melikidze}, G.~I. 1991, \mnras, 253,
  377

\bibitem[{{Kijak} \& {Gil}(1997)}]{1997MNRAS.288..631K}
{Kijak}, J., \& {Gil}, J. 1997, \mnras, 288, 631

\bibitem[{{Kramer} {et~al.}(2002){Kramer}, {Johnston}, \& {van
  Straten}}]{2002MNRAS.334..523K}
{Kramer}, M., {Johnston}, S., \& {van Straten}, W. 2002, \mnras, 334, 523

\bibitem[{{Krzeszowski} {et~al.}(2009){Krzeszowski}, {Mitra}, {Gupta}, {Kijak},
  {Gil}, \& {Acharyya}}]{2009MNRAS.393.1617K}
{Krzeszowski}, K., {Mitra}, D., {Gupta}, Y., {et~al.} 2009, \mnras, 393, 1617

\bibitem[{{Lange} {et~al.}(1998){Lange}, {Kramer}, {Wielebinski}, \&
  {Jessner}}]{1998A&A...332..111L}
{Lange}, C., {Kramer}, M., {Wielebinski}, R., \& {Jessner}, A. 1998, \aap, 332,
  111

\bibitem[{{Manchester} {et~al.}(2005){Manchester}, {Hobbs}, {Teoh}, \&
  {Hobbs}}]{2005AJ....129.1993M}
{Manchester}, R.~N., {Hobbs}, G.~B., {Teoh}, A., \& {Hobbs}, M. 2005, \aj, 129,
  1993

\bibitem[{{Manchester} {et~al.}(1975){Manchester}, {Taylor}, \&
  {Huguenin}}]{1975ApJ...196...83M}
{Manchester}, R.~N., {Taylor}, J.~H., \& {Huguenin}, G.~R. 1975, \apj, 196, 83

\bibitem[{{Melikidze} {et~al.}(2000){Melikidze}, {Gil}, \&
  {Pataraya}}]{2000ApJ...544.1081M}
{Melikidze}, G.~I., {Gil}, J.~A., \& {Pataraya}, A.~D. 2000, \apj, 544, 1081

\bibitem[{{Melikidze} {et~al.}(2014){Melikidze}, {Mitra}, \&
  {Gil}}]{2014ApJ...794..105M}
{Melikidze}, G.~I., {Mitra}, D., \& {Gil}, J. 2014, \apj, 794, 105

\bibitem[{{Michel}(1987)}]{1987ApJ...322..822M}
{Michel}, F.~C. 1987, \apj, 322, 822

\bibitem[{{Mitra} \& {Deshpande}(1999)}]{1999A&A...346..906M}
{Mitra}, D., \& {Deshpande}, A.~A. 1999, \aap, 346, 906

\bibitem[{{Mitra} {et~al.}(2009){Mitra}, {Gil}, \&
  {Melikidze}}]{2009ApJ...696L.141M}
{Mitra}, D., {Gil}, J., \& {Melikidze}, G.~I. 2009, \apjl, 696, L141

\bibitem[{{Mitra} \& {Li}(2004)}]{2004A&A...421..215M}
{Mitra}, D., \& {Li}, X.~H. 2004, \aap, 421, 215

\bibitem[{{Morris} {et~al.}(1981){Morris}, {Graham}, {Sieber}, {Bartel}, \&
  {Thomasson}}]{1981A&AS...46..421M}
{Morris}, D., {Graham}, D.~A., {Sieber}, W., {Bartel}, N., \& {Thomasson}, P.
  1981, \aaps, 46, 421

\bibitem[{{Noutsos} {et~al.}(2015){Noutsos}, {Sobey}, {Kondratiev},
  {Weltevrede}, {Verbiest}, {Karastergiou}, {Kramer}, {Kuniyoshi}, {Alexov},
  {Breton}, {Bilous}, {Cooper}, {Falcke}, {Grie{\ss}meier}, {Hassall},
  {Hessels}, {Keane}, {Os{\l}owski}, {Pilia}, {Serylak}, {Stappers}, {ter
  Veen}, {van Leeuwen}, {Zagkouris}, {Anderson}, {B{\"a}hren}, {Bell},
  {Broderick}, {Carbone}, {Cendes}, {Coenen}, {Corbel}, {Eisl{\"o}ffel},
  {Fender}, {Garsden}, {Jonker}, {Law}, {Marko}, {Masters}, {Miller-Jones},
  {Molenaar}, {Osten}, {Pietka}, {Rol}, {Rowlinson}, {Scheers}, {Spreeuw},
  {Staley}, {Stewart}, {Swinbank}, {Wijers}, {Wijnands}, {Wise}, {Zarka}, \&
  {van der Horst}}]{2015arXiv150103312N}
{Noutsos}, A., {Sobey}, C., {Kondratiev}, V.~I., {et~al.} 2015, ArXiv e-prints,
  arXiv:1501.03312

\bibitem[{{Popov} {et~al.}(2002){Popov}, {Bartel}, {Cannon}, {Novikov},
  {Kondratiev}, \& {Altunin}}]{2002A&A...396..171P}
{Popov}, M.~V., {Bartel}, N., {Cannon}, W.~H., {et~al.} 2002, \aap, 396, 171

\bibitem[{{R Core Team}(2014)}]{R2014}
{R Core Team}. 2014, R: A Language and Environment for Statistical Computing, R
  Foundation for Statistical Computing, Vienna, Austria, {\scriptsize
  \url{http://www.R-project.org/}}

\bibitem[{{Radhakrishnan} \& {Rankin}(1990)}]{1990ApJ...352..258R}
{Radhakrishnan}, V., \& {Rankin}, J.~M. 1990, \apj, 352, 258

\bibitem[{{Rankin}(1993)}]{1993ApJ...405..285R}
{Rankin}, J.~M. 1993, \apj, 405, 285

\bibitem[{{Rankin}(2007)}]{2007ApJ...664..443R}
---. 2007, \apj, 664, 443

\bibitem[{{Rickett} {et~al.}(1975){Rickett}, {Hankins}, \&
  {Cordes}}]{1975ApJ...201..425R}
{Rickett}, B.~J., {Hankins}, T.~H., \& {Cordes}, J.~M. 1975, \apj, 201, 425

\bibitem[{{Ruderman} \& {Sutherland}(1975)}]{1975ApJ...196...51R}
{Ruderman}, M.~A., \& {Sutherland}, P.~G. 1975, \apj, 196, 51

\bibitem[{{Soglasnov} {et~al.}(1983){Soglasnov}, {Popov}, \&
  {Kuzmin}}]{1983SvA....27..169S}
{Soglasnov}, V.~A., {Popov}, M.~V., \& {Kuzmin}, O.~A. 1983, \sovast, 27, 169

\bibitem[{{Soglasnov} {et~al.}(1981){Soglasnov}, {Smirnova}, {Popov}, \&
  {Kuzmin}}]{1981SvA....25..442S}
{Soglasnov}, V.~A., {Smirnova}, T.~V., {Popov}, M.~V., \& {Kuzmin}, A.~D. 1981,
  \sovast, 25, 442

\bibitem[{Wasserman(2006)}]{Wasserman2006}
Wasserman, L. 2006, All of Nonparametric Statistics (New York, USA: Springer)

\bibitem[{{Weatherall}(1998)}]{1998ApJ...506..341W}
{Weatherall}, J.~C. 1998, \apj, 506, 341

\end{thebibliography}

\begin{figure}
 \includegraphics[height=18cm,width=\textwidth]{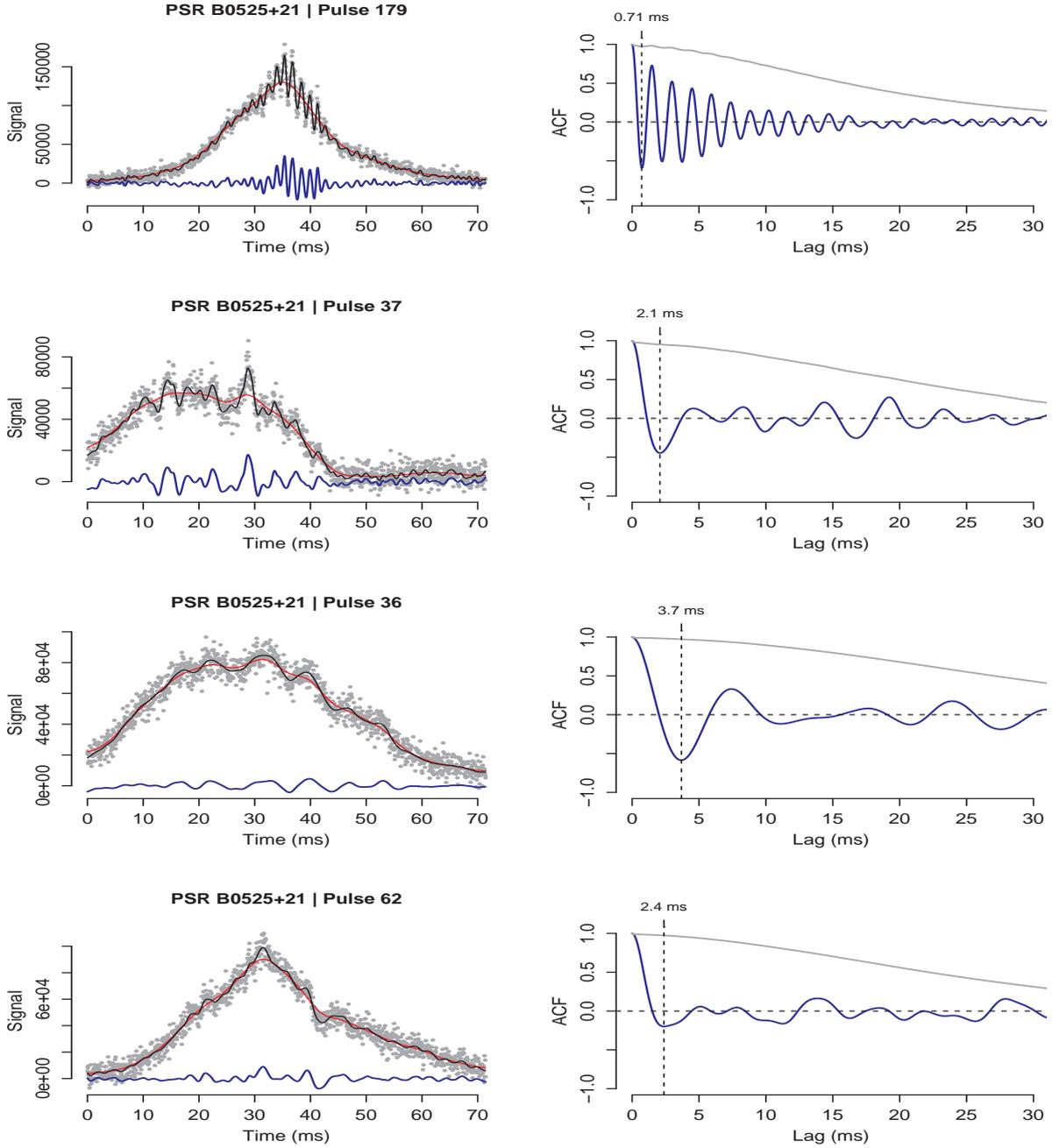}
 \caption{\label{zoo} A few illustrative subpulses from a B0525+21
 data set and their respective ACFs. Left column: gray, subpulse
 time series data; black, probable signal in the data \emph{sans} the noise;
 red, probable smooth trend in the time series; blue, probable
 microstructure feature of the signal. Right column: gray, ACF
 computed directly from noisy data, blue: ACF computed from the
 probable microstructure feature. See text for detailed description.}
\end{figure}

\begin{figure*}
\begin{center}
\begin{tabular}{@{}l@{}}
 {\mbox{\includegraphics[height=13cm,width=6cm,angle=-90.]{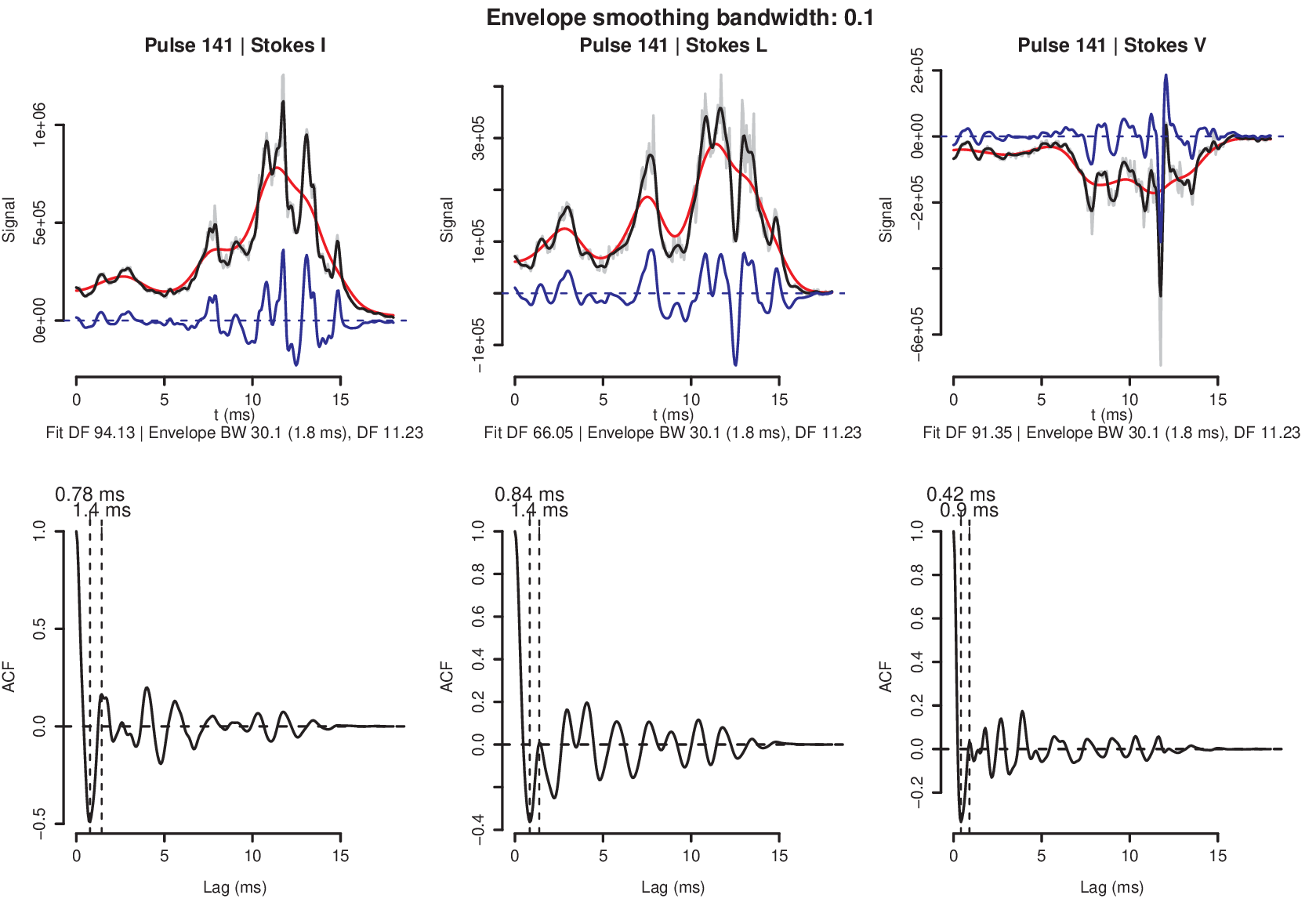}}}\\
 {\mbox{\includegraphics[height=13cm,width=6cm,angle=-90.]{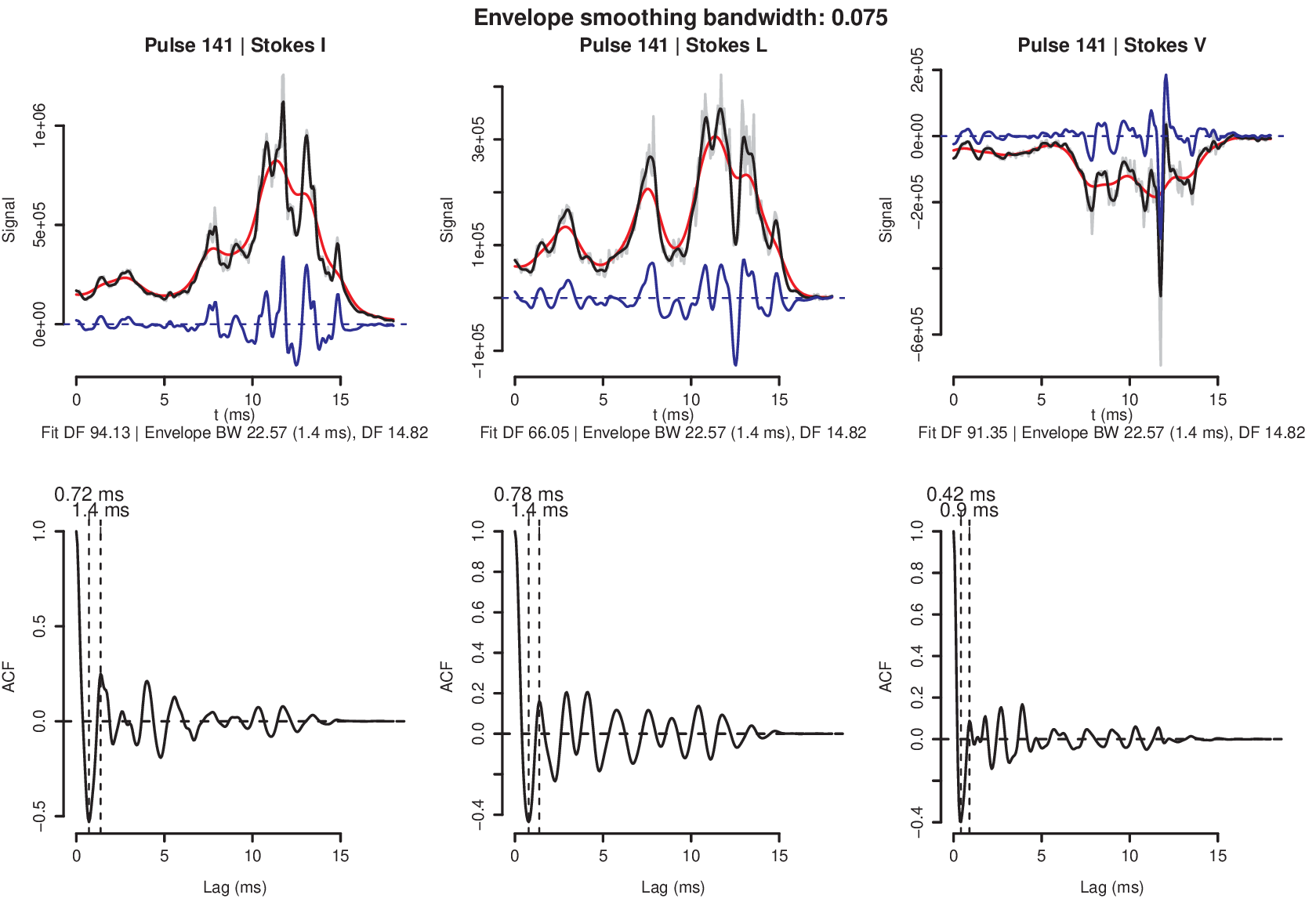}}}\\
 {\mbox{\includegraphics[height=13cm,width=6cm,angle=-90.]{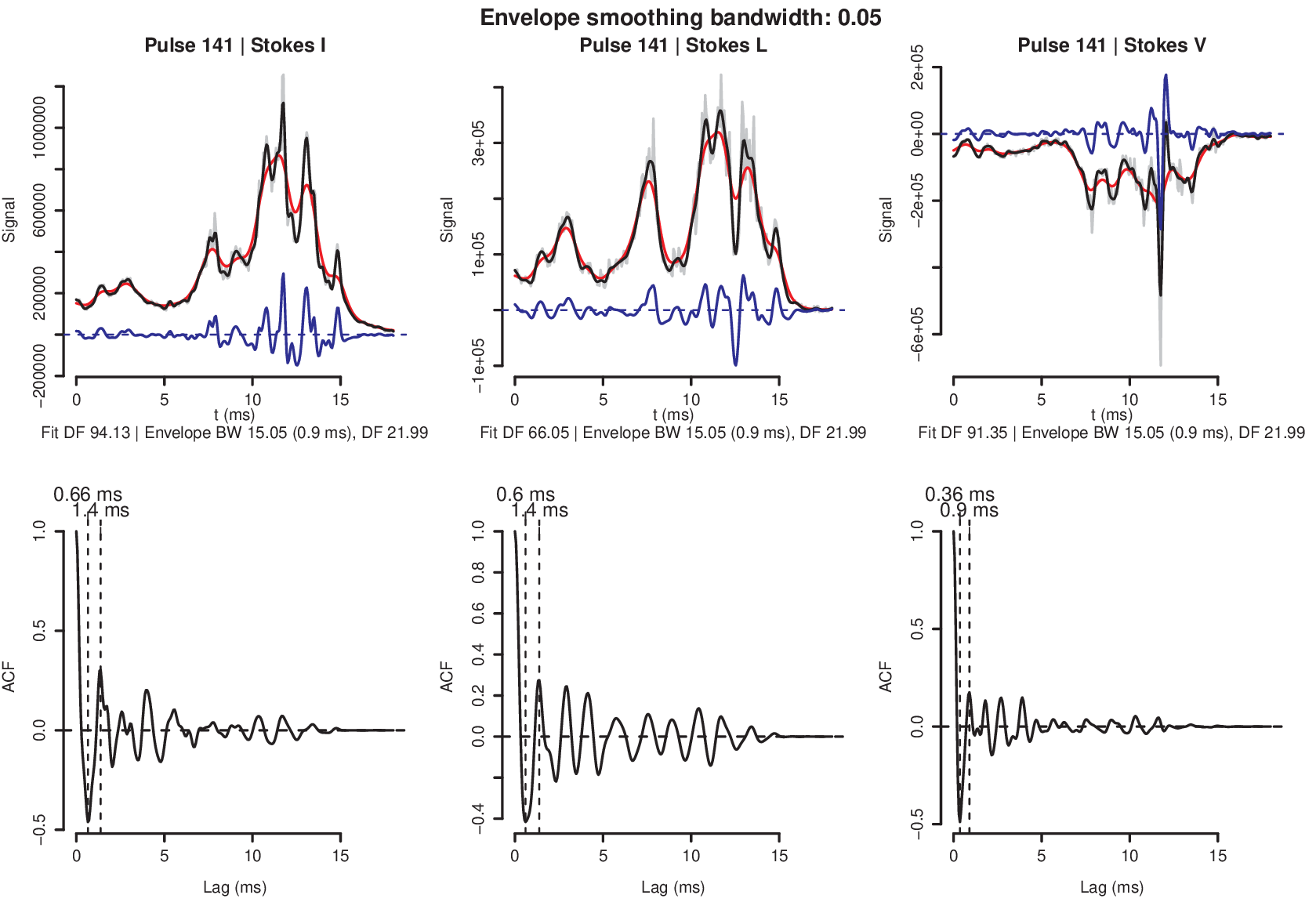}}}\\
\end{tabular}
\caption{ The analysis method for extracting the timescale of a single subpulse for the trailing 
component of PSR B1133+16 is illustrated in this figure. The plot has three columns, with 
each column corresponding to Stokes I, L and V, respectively. In each column the upper panel 
shows the subpulse, where the gray line corresponds to the subpulse data, the black line is the 
fit to the data, the red line is the envelope for a given bandwidth and the blue line is the 
microstructure signal defined as the difference between the red and black lines. The second 
panel correspond to the ACF of the microstructure (blue) signal. The timescales that we measure 
are indicated in the figure.  The three sets of plots from top to bottom are for three different 
bandwidth values 0.1, 0.075 and 0.05, respectively (see text for details). }
\label{out_r1}
\end{center}
\end{figure*}

\begin{figure*}
\begin{center}
\begin{tabular}{@{}c@{}}
 {\mbox{\includegraphics[width=6cm,height=0.7\textwidth,angle=-90.]{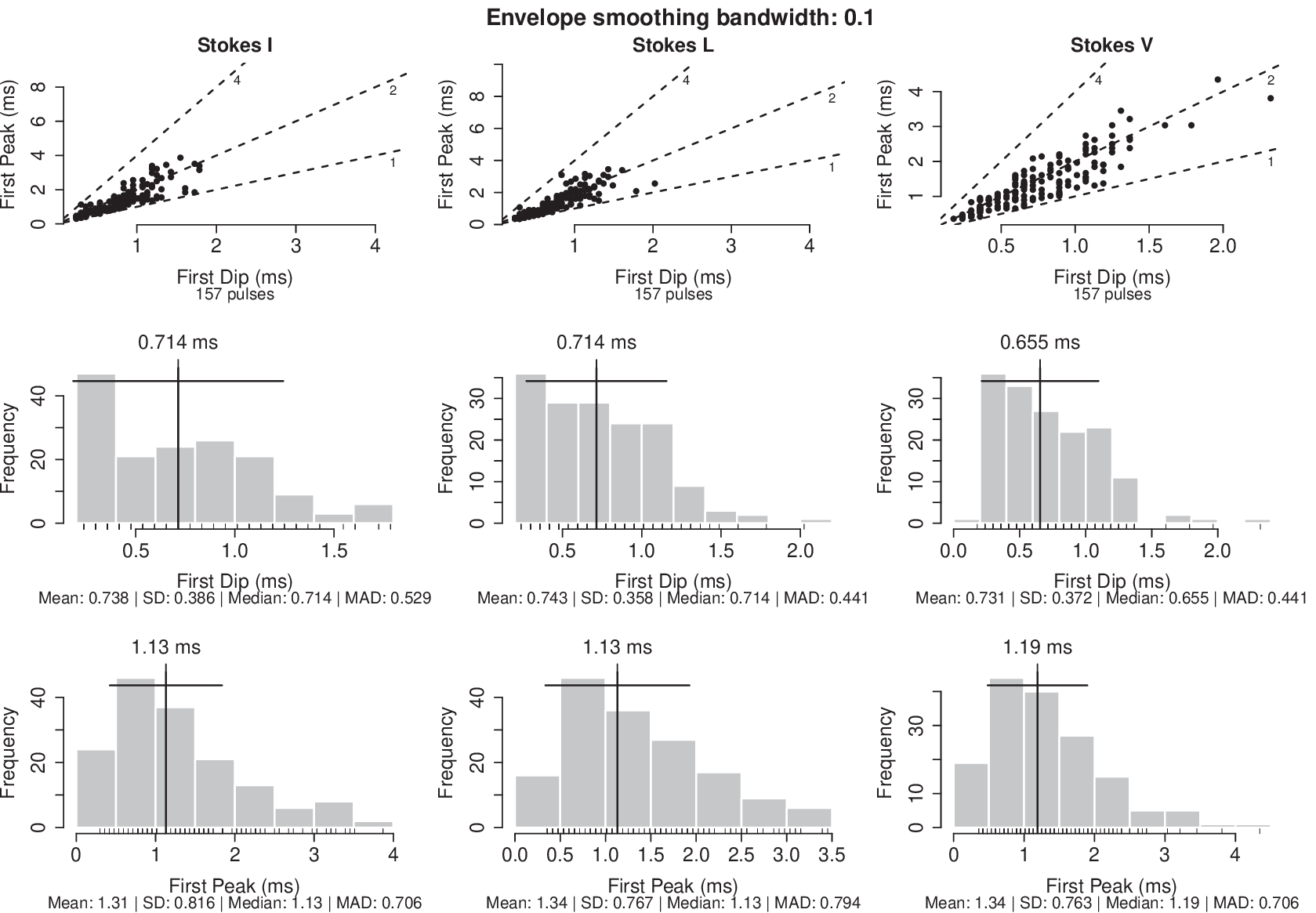}}}\\ 
 {\mbox{\includegraphics[width=6cm,height=0.7\textwidth,angle=-90.]{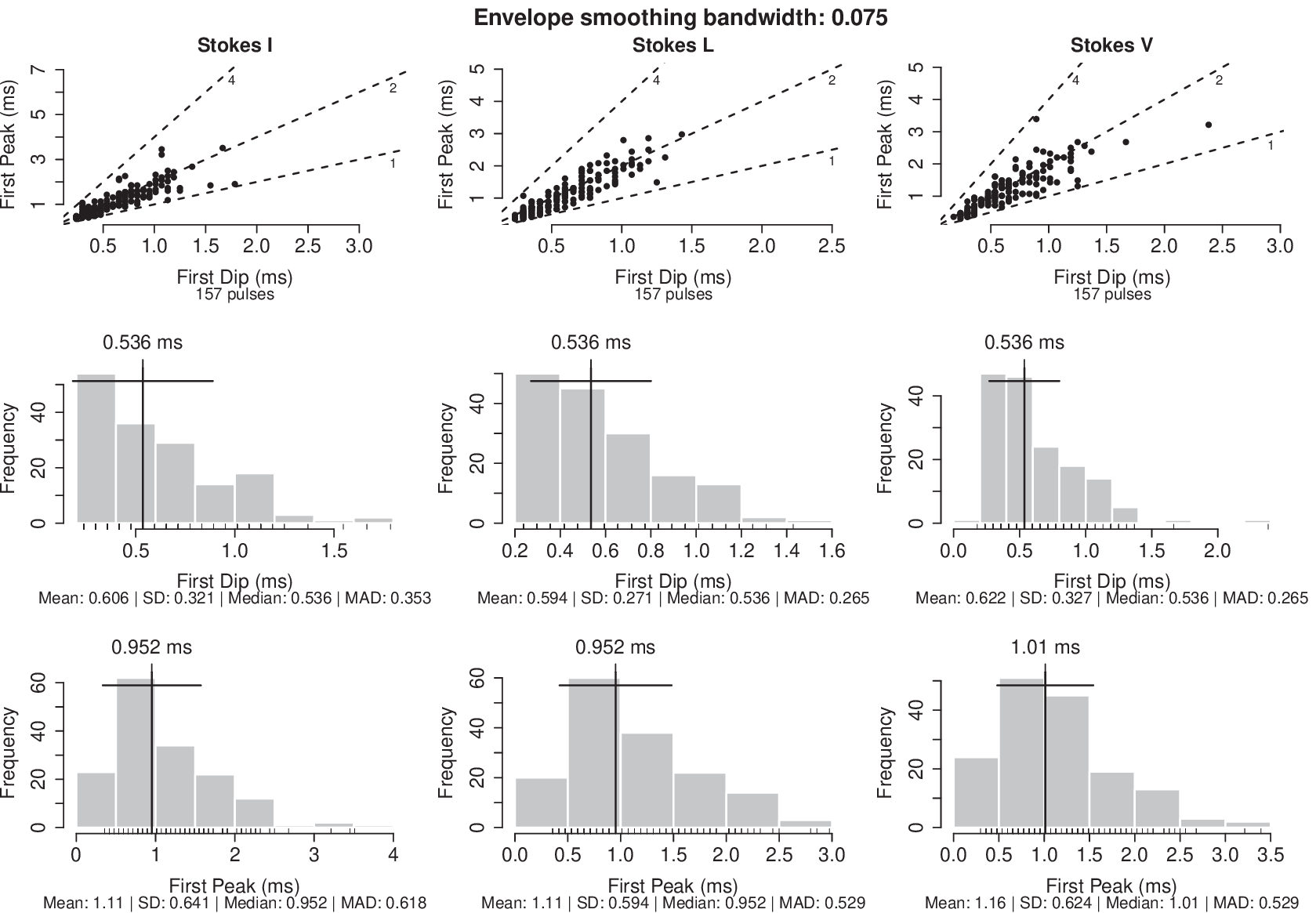}}}\\ 
 {\mbox{\includegraphics[width=6cm,height=0.7\textwidth,angle=-90.]{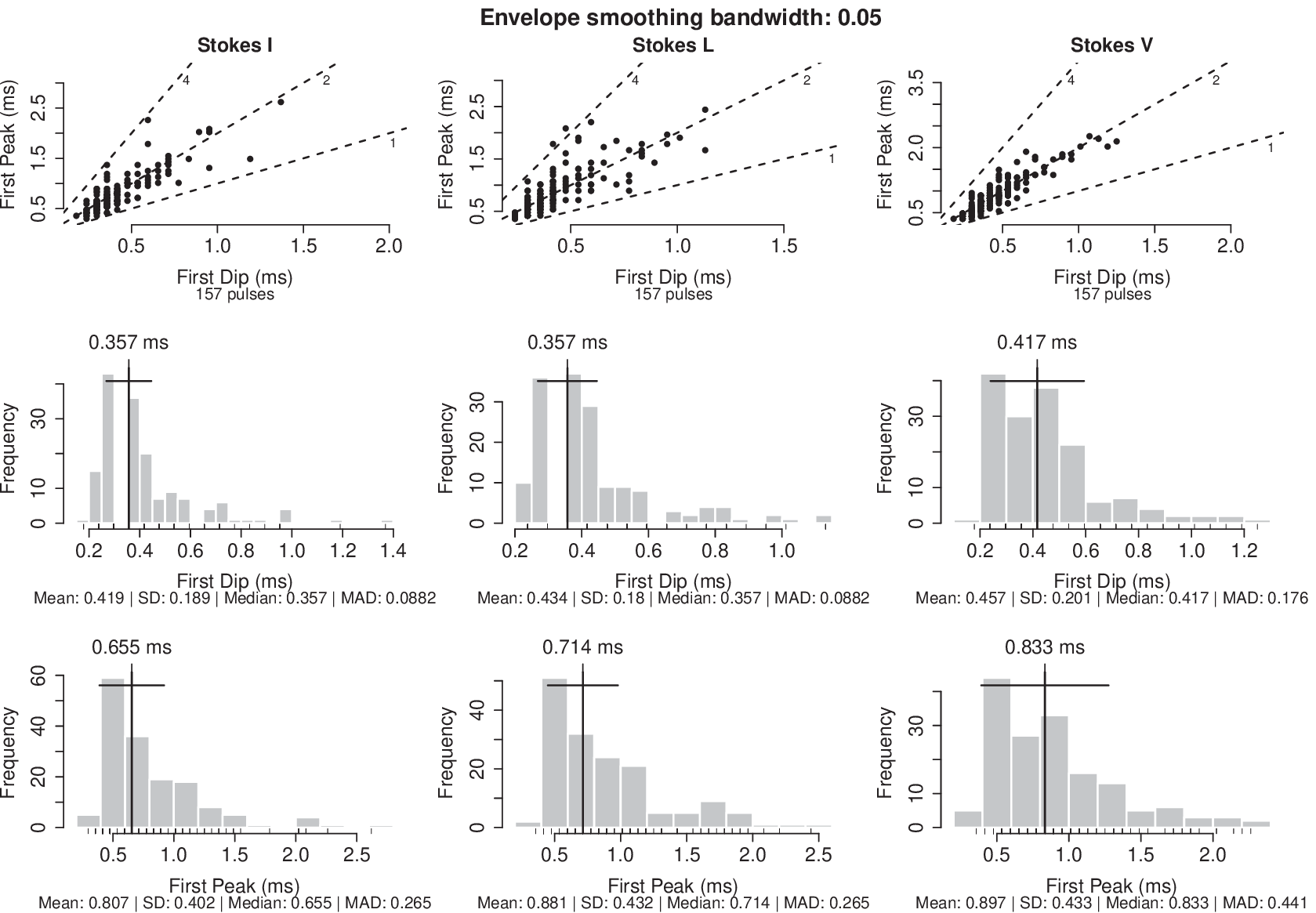}}}\\ 
\end{tabular}
\caption{The three plots above are histograms of the estimated timescales of the microstructure 
signal for the trailing component of PSR B1133+16.  
The three sets of plots correspond to three different bandwidths $h = 0.1, 0.075$ and $0.05$, respectively, from top to bottom . In each plot 
there are three columns corresponding to Stokes I, L and V from left to right, respectively.  The 
topmost panel in each plot is the relation between the first dip (as x-axis) and first peak (as y-axis)
of the ACF.  The middle panel shows distribution of the first dip in the ACF which corresponds 
to $t_{\mu}$.  The bottom panel shows the distribution of the first peak in the ACF which 
correspond to $P_{\mu}$ (see text for details). }
\label{out_r2}
\end{center}
\end{figure*}

\begin{figure}
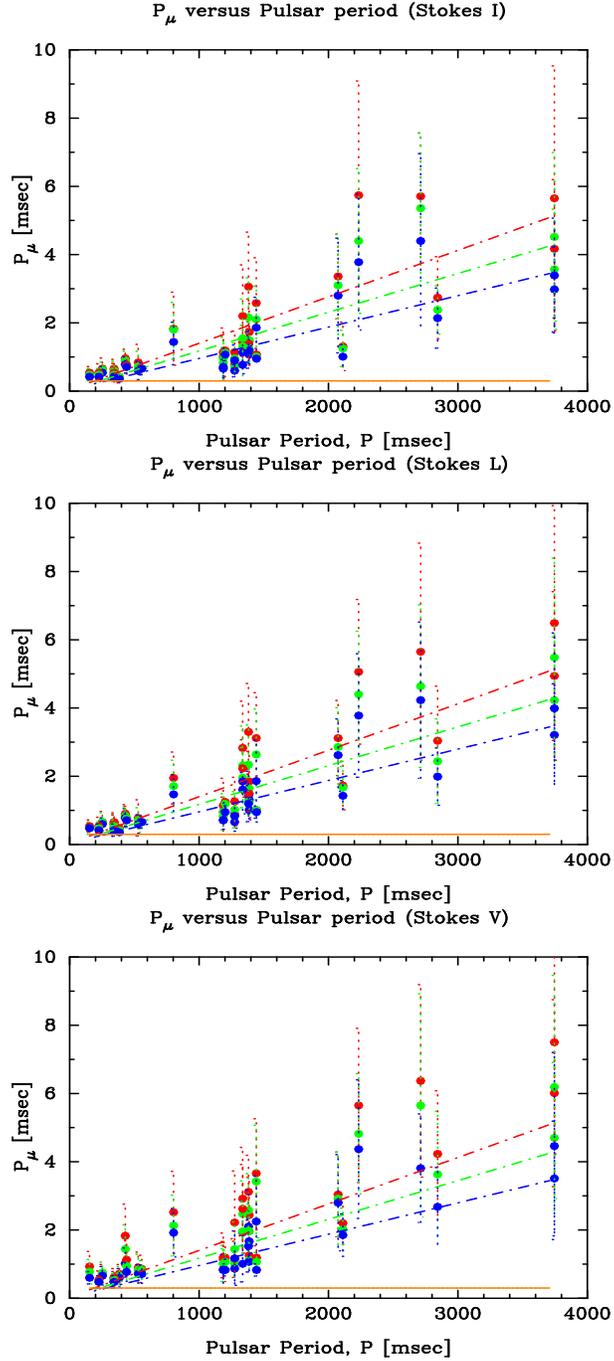

 \centerline{\includegraphics[width=6cm,height=8cm,angle=-90]{micro_per_pmu.ps}}
 \centerline{\includegraphics[width=6cm,height=8cm,angle=-90]{micro_per_pmu_l.ps}}
 \centerline{\includegraphics[width=6cm,height=8cm,angle=-90]{micro_per_pmu_v.ps}}
 \caption{\label{innerouter} The filled points and the error bars in the above plots are the median value and the 
median absolute deviation of $P_{\mu}$ distribution given in Table~\ref{tab2} 
for all the pulsars as a function
of pulsar period $P$. The three plots from top to bottom correspond
to Stokes I, L and V, respectively.  The red, green and blue points corresponds to bandwidths 
of 0.1, 0.075 and 0.05, respectively. The dashed red, green and blue line is a fit to the median $P_{\mu}$ values
for stokes I corresponding to bandwidth 0.1, 0.075 and 0.05, respectively. The same lines are 
shown for stoke L and V for reference.  
The orange line at 300 $\mu$sec along the abscissa is about five times the time resolution 
of our observations.}
\label{per_pmu}
\end{figure}

\begin{figure*}
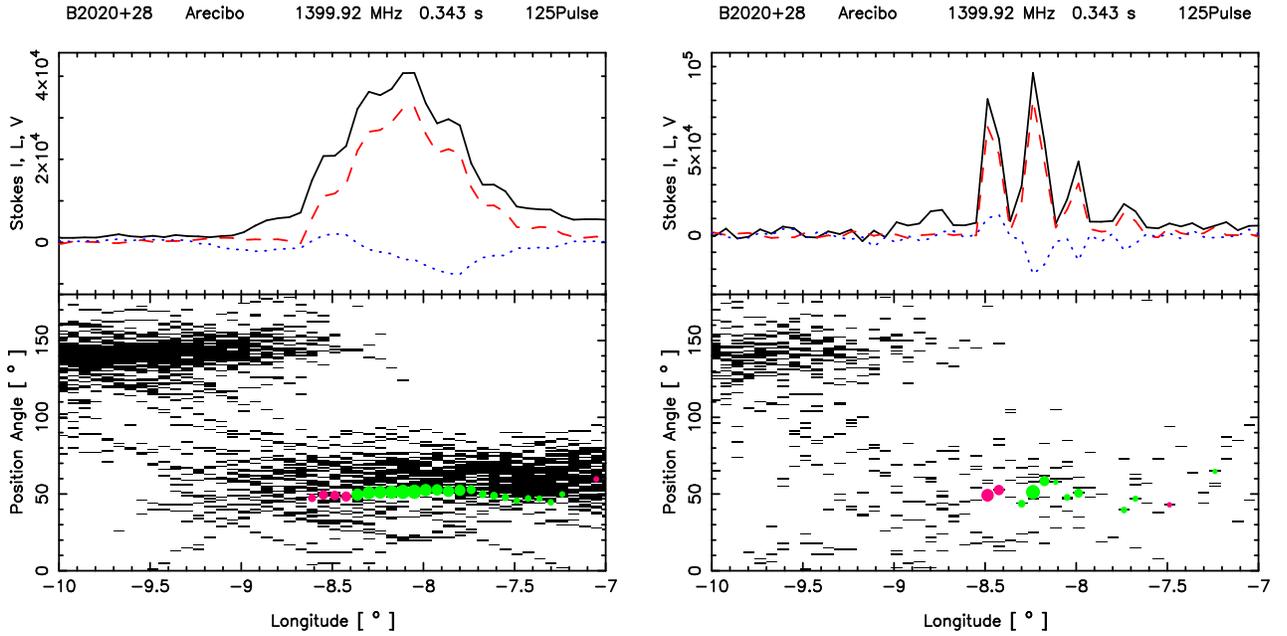

\begin{center}
\begin{tabular}{@{}lr@{}}
 {\mbox{\includegraphics[width=85mm,height=0.5\textwidth,angle=-90.]{B2020+28_p125_lowres.ps}}}& 
 {\mbox{\includegraphics[width=85mm,height=0.5\textwidth,angle=-90.]{B2020+28_p125_hires.ps}}}\\
\end{tabular}
\caption{The left plot shows a subpulse of PSR B2020+28 (plot description 
  same as given in the caption of Fig.~\ref{fig1}), 
  which has been smoothed to time resolution of 595 $\mu$sec.  Note that 
  the subpulse is close to 100\% linearly polarized and shows a sign-changing 
  circular. The corresponding right display shows the high resolution pulse with  
  time resolution 59.5 $\mu$sec.  The striking point to note there is that the 
  subpulse is split into microstructures, with the linear and circular polarization 
  following the same width and periodicity as that of the total intensity.  These 
  microstructures hence do not represent coherent curvature radiation by 
  charged bunches in vacuum (see text for details). }
\label{fig3}
\end{center}
\end{figure*}

\begin{figure*}
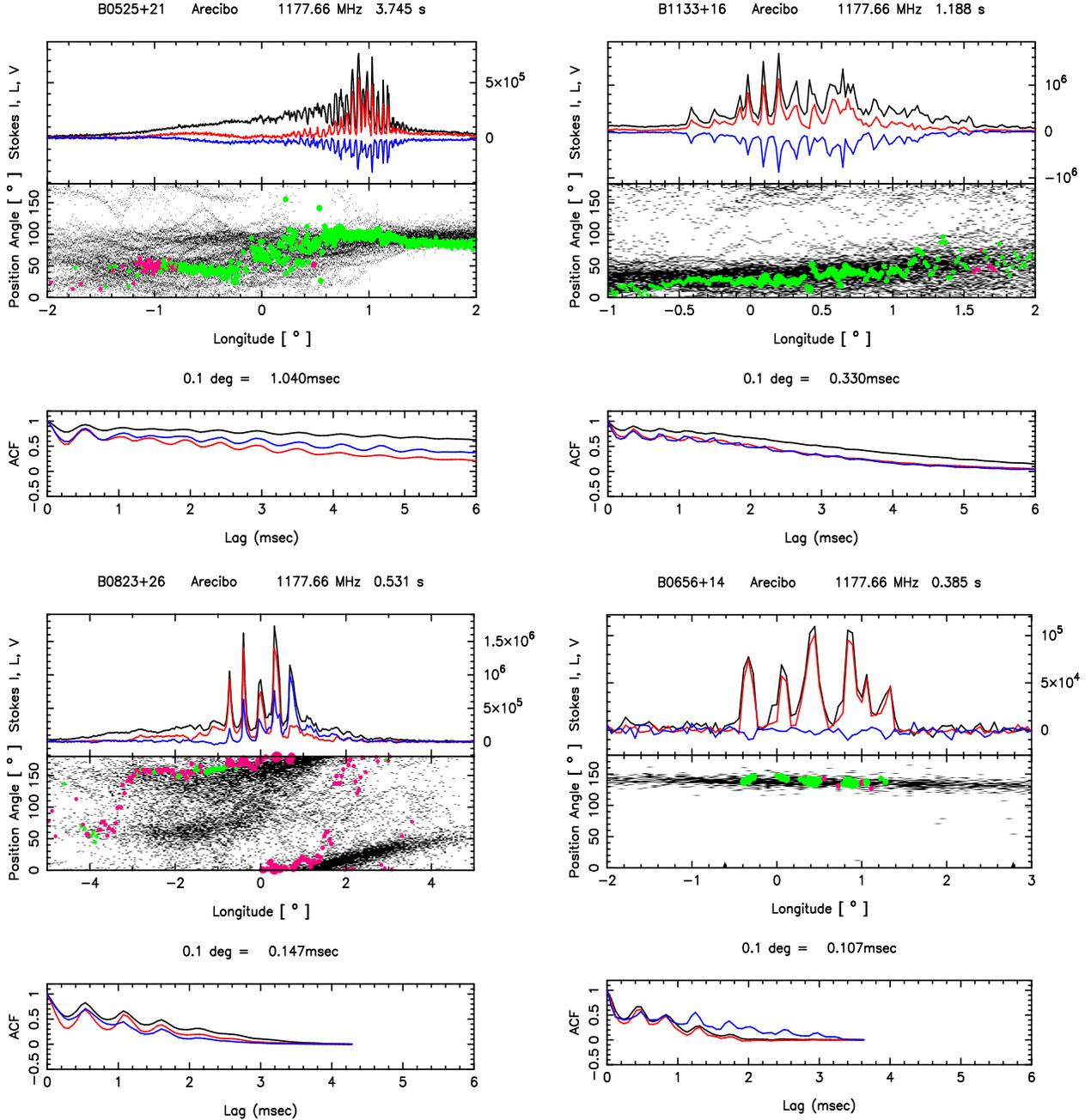

\begin{center}
\begin{tabular}{@{}lr@{}}
 {\mbox{\includegraphics[width=85mm,height=0.5\textwidth,angle=-90.]{B0525+21_VI_p60.ps}}}& 
 {\mbox{\includegraphics[width=85mm,height=0.5\textwidth,angle=-90.]{B1133+16_VI_p58.ps}}}\\
 {\mbox{\includegraphics[width=85mm,height=0.5\textwidth,angle=-90.]{B0823+26_VI_p54.ps}}}& 
 {\mbox{\includegraphics[width=85mm,height=0.5\textwidth,angle=-90.]{B0656+14_VI_p125.ps}}}\\
\end{tabular}
\caption{
  A few archetypal examples (as in Fig.~\ref{fig1}) illustrating that the circular
  polarization follows the same $P_\mu$ values as the total intensity
  and linear polarization. In each display a highly resolved subpulse is
  shown, and in the bottom panel the ACF of the Stokes parameters are
  shown. Note that in all the examples the PPA across the subpulse is 
  primarily associated with a single polarization mode. 
  Clearly the ACFs of Stokes I, L and V trace each other extremely well. }
\label{fig4}
\end{center}
\end{figure*}

\clearpage
\begin{deluxetable}{ccccccccc}
\tablecolumns{11}
\tabletypesize{\footnotesize}
\tablewidth{0pc}
\tablecaption{Observational Parameters: Column 1 to 3 in the table below lists the pulsar name, observing frequency
and the observing MJD. Column 4 gives the channel bandwidth and the total number of channels across the band.
Column 5 gives the total number of pulses observed/resolution in degrees/resolution in $\mu$sec. Column 6 and 7
gives the pulsar rotation measure and the dispersion measure (DM) used for our analysis. Column 8 and 9 
gives the pulse averaged percentage linear and circular polarization respectively. Note that pulsar name with boldface
are the ones used for microstructure timescale estimate analysis.
}
\tablehead{
\colhead{}&\colhead{}&\colhead{}&\colhead{}&\colhead{}&\colhead{}&\colhead{}&\colhead{}&\colhead{} \\
PSR  &  Band & MJD & Ch.BW (GHz)/ & length     & RM & DM & \%L &\%V \\
  Name   &  (MHz)&          & Nchan     &{res ($\degr$)/$\mu$sec} &  rad m$^{-3}$& pc/cc & & \\}
\startdata
B0301+19  & 337.668 &55283& 0.085/256 & 616/0.015/59.52 &-8.3 & 15.74 & 43 & 17 \\
B0525+21  & 351.764 &56572& 0.085/256 & 616/0.015/59.52 &-39.6 & 50.94 & 24 & -1.9  \\
J0546+2441& 351.764 &56573& 0.024/1024& 843/0.021/166.67&-19.6 & 73.81 & 6  & -3.2  \\
B1919+21  & 352.806 &55320& 0.112/512 & 1139/0.016/62.5 &-16.5 & 12.46 & 14  & 8  \\
B1929+10  & 337.668 &55320& 0.085/256 & 4639/0.094/59.52&-5.6 & 3.18& 88 & -3.6  \\
B1944+17  & 352.806 &55320& 0.112/512 & 2269/0.051/62.5 &-28.0 & 16.22& 6 & -6.6  \\
B2016+28  & 337.668 &55320& 0.085/256 &  953/0.038/59.52&-34.6 & 14.17& 23 & -0.7  \\
B2020+28  & 337.668 &55320& 0.085/256 & 4211/0.062/59.52&-74.7 & 24.64 & 44  & 0.5  \\
B2044+15  & 337.668 &55320& 0.085/256 &  960/0.019/59.52&-100.0 & 39.84& 36 &-10.3 \\
B2110+27  & 337.668 &55320& 0.085/256 & 1265/0.018/59.52&-37.0 & 25.11 & 4 & -3.3 \\
B2315+21  & 352.806 &55275& 0.112/512 &  275/0.015/59.52&-35.8 & 20.91& 32 & 3  \\
         &    &   &     &          & & & &\\
\revision{B0301+19}  & 1220.668 &55425&0.085/1024 & 1202/0.015/59.52&-8.3 & 15.74 & 36 & 8.8 \\
\revision{B0525+21}  & 1220.668 &55522&0.085/1024 &  640/0.006/59.52&-39.6 & 50.94& 45 & -13.7 \\
\revision{J0546+2441} & 1220.544 &56572&0.085/1024 &  421/0.007/59.52&-19.2  & 73.81& 35 & -13  \\
\revision{B0656+14}  & 1220.668 &55522&0.085/1024 & 1558/0.055/59.52&-23.5 & 13.997& 91 & -19.7 \\
\revision{B0751+32}  & 1220.668 &55522&0.085/1024 & 1944/0.014/59.52&-7.0 & 39.949& 29 & -26\\
\revision{B0823+26}  & 1220.668 &55522&0.085/1024 & 1130/0.040/59.52&5.9 & 19.454& 19 & 0 \\
\revision{B0834+06}  & 1220.668 &55522&0.085/1024 & 1006/0.017/59.52&23.6 & 12.889 & 15 & -4.5\\
\revision{B0919+06}  & 1220.668 &55522&0.085/1024 & 1045/0.049/59.52&29.2 & 27.271& 48 & 4.6\\
\revision{B0950+08}  & 1220.668 &55522&0.085/1024 & 2055/0.084/59.52&0.66 &  2.958 & 22 & -4.5\\
\revision{B1133+16}  & 1220.668 &55522&0.085/1024 & 2104/0.018/59.52&1.1 &  4.864 & 28 & -9\\
\revision{B1237+25}  & 1220.668 &56550&0.085/1024 &  873/0.015/59.52&-0.33 &  9.242 & 51 & 9 \\
J1503+2111      & 1220.416 &56550&0.085/1024 & 1025/0.006/59.52&16.0& 11.75 & 22 & 30\\
B1534+12        & 1663.008 &55637&0.067/512  & 15835/0.36/38.69 &10.6 & 11.614&30  & -2.9  \\
J1713+0747  &1220.668&55632&0.085/1024 & 65646/4.735/59.52&8.4& 15.9915 & 35  & -1.4 \\
J1720+2150     &  1220.544  &56550&0.086/1024 &  1021/0.013/59.52&48.0$\pm$10 & 41.4& 39 & -63 \\
\revision{B1737+13}  & 1220.668 &55632&0.085/1024 &  747/0.026/59.52&64.4 & 48.673 & 45 & -3.2 \\
\revision{J1740+1000} & 1220.668 &55632&0.085/1024 & 3893/0.139/59.52&23.8 & 23.85 & 88  & -23 \\
J1746+2245 & 1220.544 &56550&0.085/1024 & 899/0.006/59.52& 91.0$\pm$5 &52.0  & 29  & 42 \\
B1855+09 & 1220.668 &55637&0.086/1024 & 111910/4./59.52&16.4&13.295& 22 & 1.8  \\
B1901+10  & 1220.752 &56563&0.085/1024  & 1034/0.011/59.52&-105$\pm$5 &135& 36  & 35  \\
\revision{J1910+0714} &1220.752 &56563&0.085/1024  & 1024/0.007/59.52&182$\pm$20 & 124.06& 20  & -12  \\
\revision{B1910+20}  & 1220.752 &56563&0.085/1024  & 1025/0.009/59.52&148$\pm$10 &88.34 & 37  & -17 \\
\revision{B1919+21}   &1563.008 &56207& 0.672/128  & 346/0.016/59.52 &-16.5 & 12.46 &  10 & 0.6  \\
\revision{B1929+10}  & 1220.668  &55633&0.085/1024  & 2648/0.095/59.52&5.26 & 3.180& 68  & -26  \\
\revision{B1944+17}  & 1220.668 &55633&0.085/1024  & 1361/0.048/59.52&28.0 & 16.22& 22  & -14 \\
\revision{B2002+31}  & 1220.668 &56564&0.085/1024  & 1054/0.010/59.52&30.0 & 234.82& 15  & 2.6 \\
\revision{B2016+28}  & 1220.668 &55632&0.085/1024  & 1075/0.038/59.52&-34.6 & 14.172& 23 & -4  \\
\revision{B2020+28}  & 1220.668 &55632&0.085/1024  & 1747/0.062/59.52&-74.7 & 24.64 & 37  & -6.5  \\
\revision{B2034+19}  & 1220.627 &56564&0.085/1024 & 1341/0.014/59.52&-97$\pm$10 & 36.0& 24  & 5  \\
\revision{B2110+27}  & 1470.668 &55425&0.085/1024  & 1625/0.018/59.52&-37.0 & 25.113& 13  & 15 \\
B2113+14  & 1470.668 &55425&0.085/1024  & 1060/0.048/59.52&-25.0 & 56.149& 19 & -15  \\
\revision{B2315+21}  & 1470.668 &55425&0.085/1024  &  746/0.014/59.52&-35.0 & 20.906& 17 & -15  \\

\enddata
\label{tab1}
\end{deluxetable}

\clearpage
\begin{deluxetable}{lccccccc}
\tablecolumns{8}
\tablewidth{0pc}
\tablecaption{Pulsar Parameters used for microstructure study in this paper obtained
from the ATNF catalogue available in \url{http://www.atnf.csiro.au/research/pulsar/psrcat/} (\citealt{2005AJ....129.1993M}). }
\tablehead{
\colhead{}&\colhead{}&\colhead{}&\colhead{}&\colhead{}&\colhead{}&\colhead{} \\
  &PSR & Per &  $\dot{P}$  &     AGE  &   BSURF  &   $\dot{E}$ \\
  &  Name    &   sec     & ($\times 10^{-15}$)   & ($\times 10^6$ Yr)   &($\times 10^{12}$ G)       &   ($\times 10^{31}$ ergs/s) \\}
\startdata
1  &  B0301+19  &    1.387584& 1.30  &17.0 & 1.36 &1.91 \\
2  &  B0525+21  &    3.745539& 40.1  &1.48 & 12.4 &3.01 \\
3  &  J0546+2441&    2.843850& 7.65  &5.89 & 4.72 &1.31 \\
4  &  B0656+14  &    0.384891& 55.0  &0.111& 4.66 &3810 \\
5  &  B0751+32  &    1.442349& 1.08  &21.2 & 1.26 &1.42 \\
6  &  B0823+26  &    0.530661& 1.71  &4.92 & 0.964&45.2 \\
7  &  B0834+06  &    1.273768& 6.80  &2.97 & 2.98 &13.0 \\
8  &  B0919+06  &    0.430627& 13.7  &0.497& 2.46 &679 \\ 
9  &  B0950+08  &    0.253065& 0.23  &17.5 & 0.244&56.0 \\
10 &  B1133+16  &    1.187913& 3.73  &5.04 & 2.13 &8.79 \\
11 &  B1237+25  &    1.382449& 0.96  &22.8 & 1.17 &1.43 \\
12 &  J1740+1000&    0.154087& 21.5  &0.114& 1.84 &23200 \\
13 &  B1737+13  &    0.803050& 1.45  &8.77& 1.09 &11.1 \\
14 &  J1910+0714&    2.712423& 6.12  &7.02 & 4.12 &1.21 \\
15 &  B1910+20  &    2.232969& 10.2  &3.48 & 4.82 &3.61 \\
16 &  B1919+21  &    1.337302& 1.35  &15.7 & 1.36 &2.23 \\
17 &  B1929+10  &    0.226518& 1.16  &3.10 & 0.518 &393 \\
18 &  B1944+17  &    0.440618& 0.0241&290 & 0.104 &1.11 \\
19 &  B2002+31  &    2.111265& 74.6  &0.449 & 12.7 &31.3 \\
20 &  B2016+28  &    0.557953& 0.148 &59.7 & 0.291 &3.37 \\
21 &  B2020+28  &    0.343402& 1.89 &2.87 & 0.816 &185 \\
22 &  B2034+19  &    2.074377& 2.04 &16.1 & 2.08 &0.902 \\
23 &  B2110+27  &    1.202852& 2.62 &7.27 & 1.80 &5.95 \\
24 &  B2315+21  &    1.444653& 1.05 &21.9 & 1.24 &1.37 \\
\enddata
\label{tab1a}
\end{deluxetable}

\clearpage
\begin{deluxetable} {lllllllll}
\tablecolumns{8}
\tablewidth{0pc}
\tablecaption{The median $P_{\mu}$ values for microstructures for all the Stokes parameters. In column 2
the longitude range is the window that has been used for analysis and this region of the pulse can be
seen as gray regions in the average profiles given in the appendix. In column 5, 6 and 7 the median $P_{\mu}$ value is given
and the values in the parenthesis correspond to the median absolute deviation of the $P_{\mu}$ values. Column
8 gives the value of the bandwidth $h$ used for analysis and column 9 gives the number of points in the
pulse window (see text for further details).}
\tablehead{
\colhead{}&\colhead{}&\colhead{}&\colhead{}&\multicolumn{3}{l}{\underline{Median $P_{\mu}$ from ACF histogram}} & & \\
{ PSR } &Longitude&Freq. & Period        & P$_{\mu}^{I}$ &  P$_{\mu}^{L}$ &  P$_{\mu}^{V}$&  & NPT \\
  Name    &Range ($^{\circ}$)&(GHz) & (sec)         &   (msec)      &    (msec)      &    (msec)     & $h$   &    }
\startdata

B0301+19   &-2.2 --- 2.5  & 1.2  & 1.387         & 1.46 (0.57)  &   1.49 (0.57) &     1.25 (0.35)      &0.10 & 301    \\
B0301+19   &              & 1.2  & 1.387         & 1.19 (0.44)  &   1.25 (0.48) &     1.13 (0.35)      &0.075& 301    \\
B0301+19   &              & 1.2  & 1.387         & 1.07 (0.27)  &   1.01 (0.35) &     1.07 (0.35)      &0.05 & 301    \\
           &              &      &               &               &              &                     &     &        \\
B0301+19   &-11.8 --- -5.2& 1.2  & 1.387         & 1.73 (0.97)   &  1.85 (0.88) &     2.44 (1.10)     &0.10 & 427    \\
B0301+19   &              & 1.2  & 1.387         & 1.49 (0.62)   &  1.64 (0.71) &     1.96 (0.79)     &0.075& 427    \\
B0301+19   &              & 1.2  & 1.387         & 1.22 (0.40)   &  1.31 (0.40) &     1.67 (0.66)     &0.05 & 427    \\
           &              &      &               &               &              &                     &     &        \\
B0525+21   &-2.85 --- 2.86& 1.2  & 3.745         & 4.17 (2.03)  &   4.94 (2.47) &     6.01 (2.74)      &0.10 & 1001   \\
B0525+21   &              & 1.2  & 3.745         & 3.57 (1.76)  &   4.23 (1.85) &     4.70 (2.21)      &0.075& 1001   \\
B0525+21   &              & 1.2  & 3.745         & 2.98 (1.24)  &   3.21 (1.50) &     3.51 (1.68)      &0.050& 1001   \\
           &              &      &               &             &              &                     &     &        \\
B0525+21   &-14.3 --- -10.9& 1.2 & 3.745         & 5.65 (3.88)  &   6.49 (3.44) &     7.50 (3.27)      &0.10 & 1201   \\
B0525+21   &               & 1.2 & 3.745         & 4.52 (2.47)  &   5.48 (2.91) &     6.19 (3.27)      &0.075& 1201   \\
B0525+21   &               & 1.2 & 3.745         & 3.39 (1.68)  &   3.99 (2.21) &     4.46 (2.74)      &0.050& 1201   \\
           &              &      &               &             &              &                     &     &        \\
J0546+2441 &-2.13 --- 1.81& 1.2  & 2.843         & 2.74 (1.2)   &   3.04 (1.60) &     4.23 (1.85)      &0.10 & 504    \\
J0546+2441 &              & 1.2  & 2.843         & 2.38 (0.71)  &   2.44 (1.24) &     3.63 (1.85)      &0.075& 504    \\
J0546+2441 &              & 1.2  & 2.843         & 2.14 (0.88)  &   1.99 (0.84) &     2.68 (1.15)      &0.050& 504    \\
           &              &      &               &             &              &                     &     &        \\
B0656+14   &-0.61 --- 4.95& 1.2  & 0.384         & 0.42 (0.18)  &   0.48 (0.18) &     0.72 (0.27)      &0.10 & 101    \\
B0656+14   &              & 1.2  & 0.384         & 0.42 (0.09)  &   0.42 (0.09) &     0.66 (0.27)      &0.075& 101    \\
B0656+14   &              & 1.2  & 0.384         & 0.36 (0.09)  &   0.36 (0.09) &     0.60 (0.27)      &0.05 & 101    \\
           &              &      &               &             &              &                     &     &        \\
B0751+32   &-2.24 --- 2.43& 1.2  & 1.442         & 2.58 (1.33)   &   3.12 (1.33) &     3.66 (1.60)      &0.10 & 304    \\
B0751+32   &              & 1.2  & 1.442         & 2.10 (0.98)  &   2.64 (1.33) &     3.42 (1.65)      &0.075& 304    \\
B0751+32   &              & 1.2  & 1.442         & 1.86 (0.89)  &   1.86 (1.20) &     2.25 (1.33)      &0.050& 304    \\
           &              &      &               &             &              &                     &     &        \\
B0823+26   &-1.65 --- 1.41& 1.2  & 0.530         & 0.84 (0.53)  &   0.78 (0.53) &     0.90 (0.45)      &0.10 & 264    \\
B0823+26   &              & 1.2  & 0.530         & 0.72 (0.45)  &   0.72 (0.36) &     0.84 (0.45)      &0.075& 264    \\
B0823+26   &              & 1.2  & 0.530         & 0.60 (0.27)  &   0.60 (0.27) &     0.72 (0.27)      &0.050& 264    \\
           &              &      &               &             &              &                     &     &        \\
B0834+06   &-1.32 --- 1.46& 1.2  & 1.275         & 0.78 (0.35)  &   0.90 (0.46) &     1.14 (0.53)      &0.10 & 167    \\
B0834+06   &              & 1.2  & 1.275         & 0.72 (0.18)  &   0.84 (0.36) &     1.02 (0.45)      &0.075& 167    \\
B0834+06   &              & 1.2  & 1.275         & 0.60 (0.18)  &   0.66 (0.27) &     0.87 (0.49)      &0.050& 167    \\
           &              &      &               &             &              &                     &     &        \\
B0834+06   &3.7 --- 7.2   & 1.2  & 1.275         & 1.14 (0.62)  &   1.26 (0.71) &     2.22 (1.51)      &0.10 & 210    \\
B0834+06   &              & 1.2  & 1.275         & 0.96 (0.45)  &   1.02 (0.53) &     1.44 (0.80)      &0.075& 210    \\
B0834+06   &              & 1.2  & 1.275         & 0.90 (0.45)  &   0.84 (0.36) &     1.17 (0.80)      &0.050& 210    \\
           &              &      &               &             &              &                     &     &        \\
B0919+06   &-7.5 --- 4.74 & 1.2  & 0.430         & 0.96 (0.27)  &   0.90 (0.27) &     1.83 (0.93)      &0.10 & 249    \\
B0919+06   &              & 1.2  & 0.430         & 0.90 (0.27)  &   0.84 (0.27) &     1.44 (0.71)      &0.075& 249    \\
B0919+06   &              & 1.2  & 0.430         & 0.78 (0.22)  &   0.78 (0.27) &     1.02 (0.45)      &0.050& 249    \\
           &              &      &               &             &              &                     &     &        \\
B0950+08   &-12.77 --- 10.84& 1.2  & 0.253       & 0.66 (0.31)  &   0.66 (0.31) &     0.72 (0.27)      &0.10 & 279    \\
B0950+08   &                & 1.2  & 0.253       & 0.63 (0.22)  &   0.66 (0.27) &     0.75 (0.31)      &0.075& 279    \\
B0950+08   &                & 1.2  & 0.253       & 0.54 (0.18)  &   0.60 (0.18) &     0.66 (0.27)      &0.050& 279    \\
           &              &      &               &             &              &                     &     &        \\
B1133+16   &-1.47 --- 3.39  & 1.2  & 1.187       & 1.01 (0.62)  &   0.95 (0.44) &     1.13 (0.53)      &0.10 & 271    \\
B1133+16   &                & 1.2  & 1.187       & 0.83 (0.44)  &   0.83 (0.35) &     1.01 (0.44)      &0.075& 271    \\
B1133+16   &                & 1.2  & 1.187       & 0.71 (0.27)  &   0.71 (0.27) &     0.83 (0.35)      &0.050& 271    \\
           &              &      &               &             &              &                     &     &        \\
B1133+16   &3.39   --- 8.8  & 1.2  & 1.187       & 1.13 (0.71)  &   1.13 (0.80) &     1.19 (0.71)      &0.10 & 301    \\
B1133+16   &                & 1.2  & 1.187       & 0.95 (0.62)  &   0.95 (0.53) &     1.01 (0.53)      &0.075& 301    \\
B1133+16   &                & 1.2  & 1.187       & 0.66 (0.27)  &   0.71 (0.27) &     0.83 (0.44)      &0.050& 301    \\
           &              &      &               &             &              &                     &     &        \\
B1237+25   &-2.07 --- 4.3 & 1.2  & 1.382         & 3.06 (1.60)  &   3.30 (1.42) &     3.12 (1.07)      &0.01 & 401    \\
B1237+25   &              & 1.2  & 1.382         & 2.16 (1.16)  &   2.34 (1.16) &     2.58 (1.07)      &0.075& 401    \\
B1237+25   &              & 1.2  & 1.382         & 1.38 (0.53)  &   1.44 (0.71) &     2.10 (1.07)      &0.050& 401    \\
           &              &      &               &             &              &                     &     &        \\
B1237+25   &8.77 --- 11.87& 1.2  & 1.382         & 1.38 (0.45)  &   1.50 (0.71) &     1.98 (0.98)      &0.01 & 201    \\
B1237+25   &              & 1.2  & 1.382         & 1.26 (0.45)  &   1.26 (0.45) &     1.98 (0.98)      &0.075& 201    \\
B1237+25   &              & 1.2  & 1.382         & 1.20 (0.53)  &   1.20 (0.53) &     1.53 (0.76)      &0.050& 201    \\
           &              &      &               &              &              &                     &     &        \\
J1740+1000 &-4.87 --- 10.7& 1.2  & 0.154         & 0.54 (0.18)  &   0.54 (0.18) &     0.93 (0.44)      &0.10 & 113    \\
J1740+1000 &              & 1.2  & 0.154         & 0.48 (0.18)  &   0.48 (0.18) &     0.78 (0.36)      &0.075& 113    \\
J1740+1000 &              & 1.2  & 0.154         & 0.42 (0.09)  &   0.48 (0.09) &     0.60 (0.18)      &0.050& 113    \\
           &              &      &               &              &              &                     &     &        \\
B1737+13   &-7.94 --- 5.41& 1.2  & 0.803         & 1.83 (1.07)  &   1.95 (0.76) &     2.52 (1.20)      &0.10 & 501    \\
B1737+13   &              & 1.2  & 0.803         & 1.80 (0.80)  &   1.71 (0.76) &     2.13 (0.89)      &0.075& 501    \\
B1737+13   &              & 1.2  & 0.803         & 1.44 (0.58)  &   1.47 (0.58) &     1.92 (0.71)      &0.050& 501    \\
           &              &      &               &              &              &                     &     &        \\
J1910+0714 &-1.41 --- 3.91& 1.2  & 2.712         &  5.71 (1.85) &   5.65 (3.18)&     6.37 (2.82)      &0.10 & 675    \\
J1910+0714 &              & 1.2  & 2.712         &  5.36 (2.21) &   4.64 (2.38)&     5.65 (3.27)      &0.075& 675    \\
J1910+0714 &              & 1.2  & 2.712         &  4.40 (2.56) &   4.23 (2.29)&     3.81 (1.59)      &0.050& 675    \\
           &              &      &               &             &              &                     &     &        \\
B1910+20   &-1.21 --- 6.46& 1.2  & 2.233         & 5.74 (3.35) &   5.06 (2.12)&     5.65 (2.03)      &0.10 & 801    \\
B1910+20   &              & 1.2  & 2.233         & 4.40 (2.12) &   4.40 (1.85)&     4.82 (1.76)      &0.075& 801    \\
B1910+20   &              & 1.2  & 2.233         & 3.78 (1.99) &   3.78 (1.81)&     4.37 (2.03)      &0.050& 801    \\
           &              &      &               &             &              &                     &     &        \\
B1919+21   &-2.45 --- 3.10& 1.5  & 1.337         &  1.43 (0.71) &   2.83 (1.37)&     2.92 (1.50)     &0.10 & 348    \\
B1919+21   &              & 1.5  & 1.337         &  1.31 (0.62) &   2.32 (1.15)&     2.47 (0.97)     &0.075& 348    \\
B1919+21   &              & 1.5  & 1.337         &  1.13 (0.35) &   1.61 (0.71)&     1.93 (0.93)     &0.050& 348    \\
           &              &      &               &             &              &                     &     &        \\
B1919+21   &5.28 --- 8.48 & 1.5  & 1.337         &  2.20 (1.50) &   2.23 (0.84)&     2.62 (1.15)     &0.10 & 201    \\
B1919+21   &              & 1.5  & 1.337         &  1.55 (1.15) &   1.96 (0.88)&     1.96 (0.88)     &0.075& 201    \\
B1919+21   &              & 1.5  & 1.337         &  0.77 (0.26) &   1.85 (0.26)&     1.01 (0.53)     &0.050& 201    \\
           &              &      &               &             &              &                     &     &        \\
B1929+10   &-11.1 --- 7.84& 1.2  & 0.226         &  0.54 (0.26) &   0.54 (0.26)&     0.54 (0.26)      &0.10 & 201    \\
B1929+10   &              & 1.2  & 0.226         &  0.48 (0.18) &   0.48 (0.18)&     0.48 (0.18)      &0.075& 201    \\
B1929+10   &              & 1.2  & 0.226         &  0.42 (0.09) &   0.42 (0.18)&     0.48 (0.18)      &0.050& 201    \\
           &              &      &               &             &              &                     &     &        \\
B1944+17   &-9.67 --- 4.91& 1.2  & 0.440         &  0.77 (0.27) &   0.77 (0.35)&     1.13 (0.57)      &0.10 & 301    \\
B1944+17   &              & 1.2  & 0.440         &  0.71 (0.18) &   0.77 (0.27)&     0.95 (0.44)      &0.075& 301    \\
B1944+17   &              & 1.2  & 0.440         &  0.71 (0.18) &   0.71 (0.18)&     0.77 (0.27)      &0.050& 301    \\
           &              &      &               &             &              &                     &     &        \\
B2002+31   &-2.48 --- 2.01& 1.2  & 2.111         &  1.31 (0.71) &   1.73 (0.71)&     2.20 (0.62)      &0.10 & 445    \\
B2002+31   &              & 1.2  & 2.111         &  1.25 (0.53) &   1.67 (0.62)&     2.02 (0.62)      &0.075& 445    \\
B2002+31   &              & 1.2  & 2.111         &  1.01 (0.35) &   1.43 (0.44)&     1.85 (0.62)      &0.050& 445    \\
           &              &      &               &             &              &                     &     &        \\
B2016+28   &-5.34 --- 4.45& 1.2  & 0.557         &  0.66 (0.18) &   0.66 (0.18)&     0.86 (0.40)      &0.10 & 256    \\
B2016+28   &              & 1.2  & 0.557         &  0.66 (0.18) &   0.66 (0.18)&     0.83 (0.27)      &0.075& 256    \\
B2016+28   &              & 1.2  & 0.557         &  0.66 (0.18) &   0.66 (0.18)&     0.71 (0.27)      &0.050& 256    \\
           &              &      &               &             &              &                     &     &        \\
B2020+28   &-11.27 --- -5.03& 1.2& 0.343         &  0.66 (0.27) &   0.66 (0.27)&     0.60 (0.18)      &0.10 & 101    \\
B2020+28   &                & 1.2& 0.343         &  0.60 (0.27) &   0.60 (0.27)&     0.54 (0.18)      &0.075& 101    \\
B2020+28   &                & 1.2& 0.343         &  0.48 (0.18) &   0.48 (0.18)&     0.48 (0.18)      &0.050& 101    \\
           &              &      &               &             &              &                     &     &        \\
B2020+28   &-3.78 --- 1.21& 1.2  & 0.343         &  0.54 (0.27) &   0.60 (0.27)&     0.60 (0.27)      &0.10 &  98    \\
B2020+28   &              & 1.2  & 0.343         &  0.48 (0.18) &   0.54 (0.18)&     0.54 (0.18)      &0.075&  98    \\
B2020+28   &              & 1.2  & 0.343         &  0.42 (0.09) &   0.42 (0.18)&     0.48 (0.18)      &0.050&  98    \\
           &              &      &               &             &              &                     &     &        \\
B2034+19   &-5.25 --- -1.49& 1.2 & 2.074         &  3.36 (1.24) &   3.12 (1.10)&     3.04 (1.24)      &0.10 & 361    \\
B2034+19   &               & 1.2 & 2.074         &  3.10 (1.50) &   2.86 (1.15)&     2.92 (1.37)      &0.075& 361    \\
B2034+19   &               & 1.2 & 2.074         &  2.80 (1.68) &   2.62 (1.06)&     2.80 (1.41)      &0.050& 361    \\
           &              &      &               &             &              &                     &     &        \\
B2110+27   &-1.3 --- 2.26  & 1.47  & 1.202        &  1.19 (0.09)&   1.25 (0.09)&     1.19 (0.44)      &0.10 & 201    \\
B2110+27   &               & 1.47  & 1.202        &  1.13 (0.09)&   1.19 (0.09)&     1.07 (0.35)      &0.075& 201    \\
B2110+27   &               & 1.47  & 1.202        &  1.07 (0.18)&   0.95 (0.27)&     0.83 (0.18)      &0.050& 201    \\
           &              &      &               &             &              &                     &     &        \\
B2315+21   &-1.79 --- 1.17 & 1.47  & 1.444        &  1.07 (0.09)&   1.01 (0.09)&     1.19 (0.35)     &0.10  & 201   \\
B2315+21   &               & 1.47  & 1.444        &  1.01 (0.09)&   1.01 (0.09)&     1.07 (0.27)     &0.075 & 201   \\
B2315+21   &               & 1.47  & 1.444        &  0.95 (0.09)&   0.95 (0.18)&     0.83 (0.18)     &0.050 & 201   \\
\enddata
\label{tab2}
\end{deluxetable}

\clearpage
\appendix
\setcounter{figure}{0}
\renewcommand{\thefigure}{A\arabic{figure}}
\renewcommand{\thetable}{A\arabic{table}}
\setcounter{table}{0}
\renewcommand{\thefootnote}{A\arabic{footnote}}
\setcounter{footnote}{0}

\section{Methods}

\subsection{Models for data and noise}
\label{a_data}
The measured signal $Y_i$ at time $t_i$ is modeled as 
\begin{equation} Y_i = f(t_i) + \epsilon_i,
\label{data.model} 
\end{equation} 
where $f$ embodies the true but unknown variation of the subpulse
signal with time, and $\epsilon_i$ represents the noise in the data.
The $N$ time points $t_i$ where signal is
measured are assumed to be equidistant; \ie of the form $t_i =
i\delta$ where $i = 0, 1, \ldots, N-1$.  ($\delta = 59.5 \mu$s is the
time resolution for all the data sets analyzed in this work).  Without
loss of generality, we take $\delta=1$ (which amounts to redefining
the time unit).
The noise variables $\epsilon_i$ are assumed to be independent with mean 0 and a (constant) variance $\sigma^2$.
Our present analysis does not require an estimate of the noise variance explicitly
(except for the purpose of choosing high-S/N pulses; see Sec.\ \ref{sec2}).
We have modeled the true signal $f$
underlying the subpulse data as 
\begin{equation}
 f(t) = e(t) + m(t), 
 \label{fet}
\end{equation}
where $e$ and $m$ respectively represent a smooth slow-varying envelope and
faster-scale quasiperiodic modulations corresponding to
microstructure.  The methodology in Sec.\ \ref{a_fit} is used to
estimate $f$ optimally, whereas $e$ is estimated using the method in
Sec.\ \ref{a_envelope}.  The difference $f(t)-e(t)$ is taken to be the
estimated microstructure.  Fig.~\ref{zoo} shows examples of such
fit-envelope-microstructure decomposition for a few subpulses.
The model~\ref{data.model} is inspired by the observation based on Fig.~\ref{zoo} 
that the ACF is dominated by the envelope feature which should be removed 
from the data to extract the microstructure.

\subsection{Outliers}
\label{a_outliers}

Radio-frequency interference (RFI) over timescales that are much
smaller than the pulse period may introduce isolated outliers in the
subpulse time series which may have detrimental effect on the
downstream analysis.
For example, outliers may cause fluctuations in the fit that may have an effect on the ACF used for timescale identification.
Therefore, before engaging into any further
analysis, we first assess the extent of contamination in a data set
(through visual inspection of subpulses) and, if deemed necessary,
we identify (and ``curate'') potential outliers in all the subpulse
time series in the data set heuristically as explained below.
In this work, we have used this approach because of its simplicity.
However, we note that a better alternative to this heuristic approach could
be the use of robust nonparametric regression methods
\citep{Haerdle1992,Efromovich1999} that deal with outliers in the
regression context in a unified manner, or outlier identification
methods with well-established formal properties.

\begin{figure}
 \includegraphics[height=\textwidth,angle=-90]{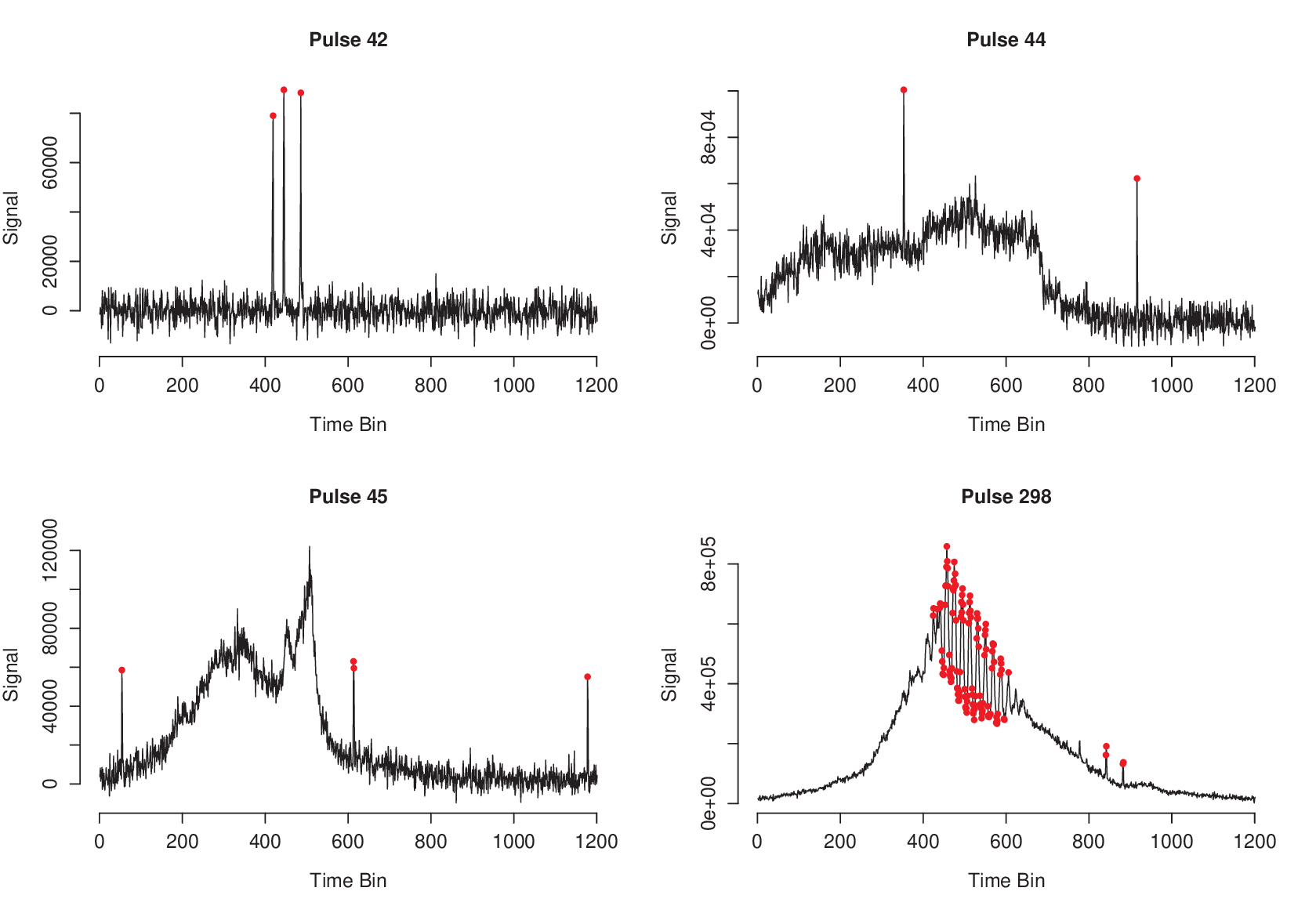}
 \caption{\label{outliers} Subpulses 42, 44, 45, and 298 from a PSR B0525+21 data set illustrating the behavior of our heuristic outlier identification method. See Sec.\ \ref{a_outliers} for description.}
\end{figure}

Fig.~\ref{outliers} shows examples of subpulses (\#42, \#44, \#45) with clear outliers from a PSR B0525+21 data set.
In fact, this was the only data set in the present study which showed clear outliers.
Further, for this data set, it was known that these outliers were caused by radio communication from a satellite in the field of view.
Fig.\ \ref{outliers} also shows a subpulse (\#298) from the same data set which illustrates a failure of our heuristic method for outlier identification.
This subpulse has strong and highly ringing microstructure which results into too many data points in the ringing part being wrongly tagged as potential outliers.
Since we expect at most a few isolated outliers in any subpulse (say, less than 1\% of the subpulse length),
such over-identification can be easily detected.
Outlier correction, as described below, is not applied to such subpulses.

\paragraph{Identifying potential outliers.}  
When the data contain an unknown trend $f$, it is necessary to identify
a potential outlier with respect to data values in its local
neighborhood.  Therefore, we first obtain a pilot fit $\widehat{f}_o$ through kernel regression \citep{Wasserman2006,Haerdle1992}.
The smoothing bandwidth for this pilot fit may be taken as a constant $b$ times the optimal smoothing bandwidth obtained via cross-validation;
slight oversmoothing ($b>1$) is expected to help make the fit $\widehat{f}_o$ somewhat robust against outliers.
Any data point $i$ with a relatively large residual $r_i = Y_i - \widehat{f}_o(t_i)$ is a potential outlier:
Specifically, following the standard prescription \citep{FHI1989}, the
$i$th data point is tagged as a potential outlier if $r_i > r_{75} +
a \times ( r_{75} - r_{25} )$ or $r_i < r_{25} - a \times ( r_{75}
- r_{25} )$, where $r_{75}$ and $r_{25}$ are respectively the 75 and
25 percentile points for the residual vector ($r_{75} - r_{25}$ is
therefore the interquartile range). The value of the multiplier $a$
is taken to be 3 instead of the conventional 1.5 to make outlier identification more conservative.

\paragraph{Imputing potential outliers back into data.}
We replace a potential isolated outlier value $Y_i$ by the median of
signal values in a small local neighborhood of $t_i$.  In this work,
we have taken the size of the local neighborhood to be 5 points on either
side of $t_i$.  That is, we replace a potential outlier value $Y_i$
with the median of $Y_{i-5}, Y_{i-4}, Y_{i-3}, Y_{i-2}, Y_{i-1},
Y_{i+1}, Y_{i+2}, Y_{i+3}, Y_{i+4}, Y_{i+5}$, with data boundaries
handled by truncation.

\subsection{Model-independent fit to subpulse time series data}
\label{a_fit}

The immense variability in the pulse shapes suggests that the
separation of probable signal from noise is best done using a
nonparametric regression method that only makes mild assumptions about
the true but unknown regression function without pre-specifying its
mathematical form.  In particular, we use smoothing spline regression
\citep{GS1994} which may be seen as a penalized least-squares method
with a penalty on the roughness of the fit.  Optimal value for the
smoothing parameter is chosen using the standard method of
(generalized) cross-validation.  The smoothness of the resultant fit
can be expressed in terms of \emph{degrees of freedom} (DF).  DF can
be interpreted as the effective number of free parameters of the fit.
Specifically, a fit with greater fluctuation has larger DF, while a
smoother fit has smaller DF.

\subsection{Estimating the pulse envelope}
\label{a_envelope}

We use the Nadaraya-Watson kernel smoother
\citep{Wasserman2006,Haerdle1992} to estimate the envelope of a
subpulse by judiciously \emph{oversmoothing} the data.  Through
trial-and-error, we found that the smoothing bandwidth $H \approx 0.075 \times
N$ is more or less reasonable for all the pulsar data sets we
analyzed.  However, considering that the smoothing bandwidth for the
envelope is perhaps the most critical parameter of our methodology,
and that there does not appear to be any principled way of arriving at
a prescription for it, we report results for a number of values of $h
\equiv H / N$ over a plausible range; the rational for this range is
discussed in Sec.\ \ref{results}.

\section{Plots for average profiles}
\label{a_avprof}

The average profiles and polarization position angle (PPA) histograms
for all the pulsars used for microstructure analysis given as boldface in Table.~\ref{tab1} are presented. 

\begin{figure*}
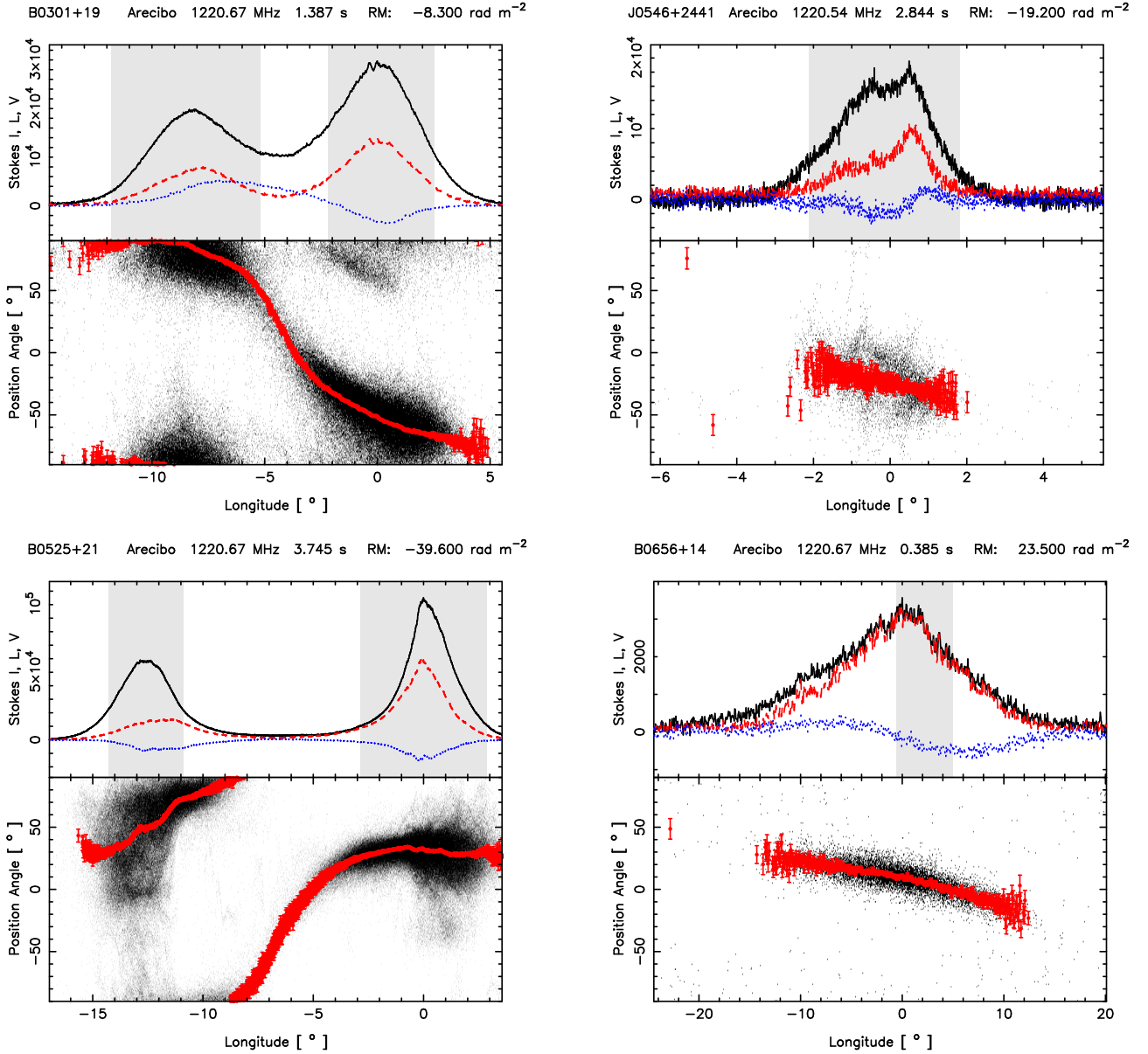

\begin{center}
\begin{tabular}{@{}lr@{}}
 {\mbox{\includegraphics[width=78mm,angle=-90.]{srB0301+19.55425l_a.ps}}}& \ \ \ \ \ \
 {\mbox{\includegraphics[width=78mm,angle=-90.]{srJ0546+2441.56572l_a.ps}}}\\
 {\mbox{\includegraphics[width=78mm,angle=-90.]{srB0525+21.55522l_a.ps}}}& \ \ \ \ \ \
 {\mbox{\includegraphics[width=78mm,angle=-90.]{srB0656+14.55522l_a.ps}}}\\
\end{tabular}
\caption{PPA histograms as in Fig.~\ref{figA1} for pulsars B0301+19, J0546+2441, B0525+21 and B0656+14
observed with time resolutions of 59.5$\mu$sec.
where the instrument and band is indicated above each plot. The respective
upper panels give the total power (black), total linear (red) and circular polarization LH-RH (blue).
The lower panels give the polarization-angle density as black dots and the 
average PPA is plotted as red points.}

\label{figA1}
\end{center}
\end{figure*}

\begin{figure*}
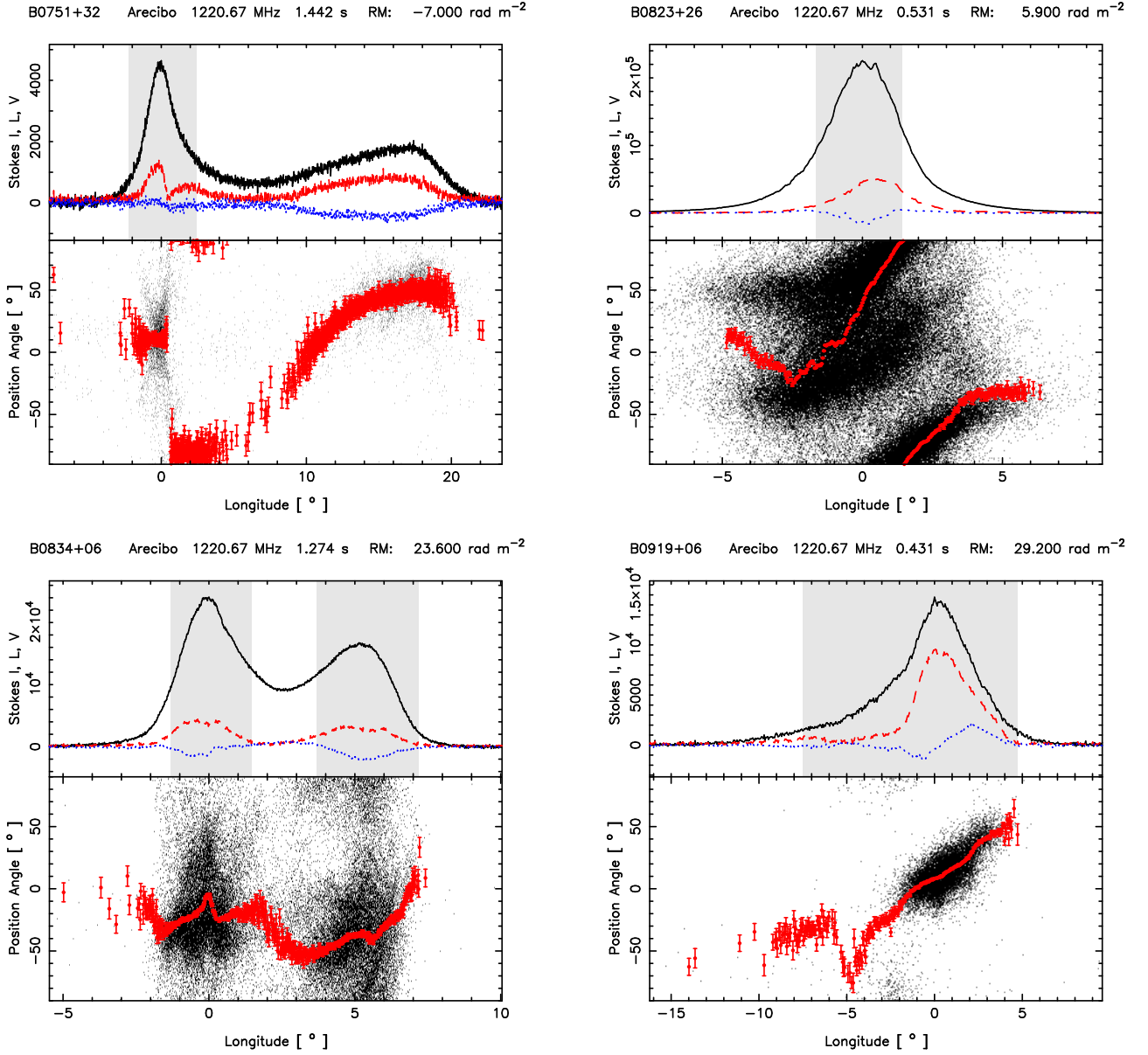

\begin{center}
\begin{tabular}{@{}lr@{}}
 {\mbox{\includegraphics[width=78mm,angle=-90.]{srB0751+32.55522l_a.ps}}}& \ \ \ \ \ \
 {\mbox{\includegraphics[width=78mm,angle=-90.]{srB0823+26.55522l_a.ps}}}\\
 {\mbox{\includegraphics[width=78mm,angle=-90.]{srB0834+06.55522l_a.ps}}}& \ \ \ \ \ \
 {\mbox{\includegraphics[width=78mm,angle=-90.]{srB0919+06.55522l_a.ps}}}\\
\end{tabular}
\caption{PPA histograms as in Fig.~\ref{figA1} for pulsars B0751+32, B0823+26, B0834+06 and B0919+06
observed with time resolutions of 59.5$\mu$sec.}
\label{figA2}
\end{center}
\end{figure*}

\begin{figure*}
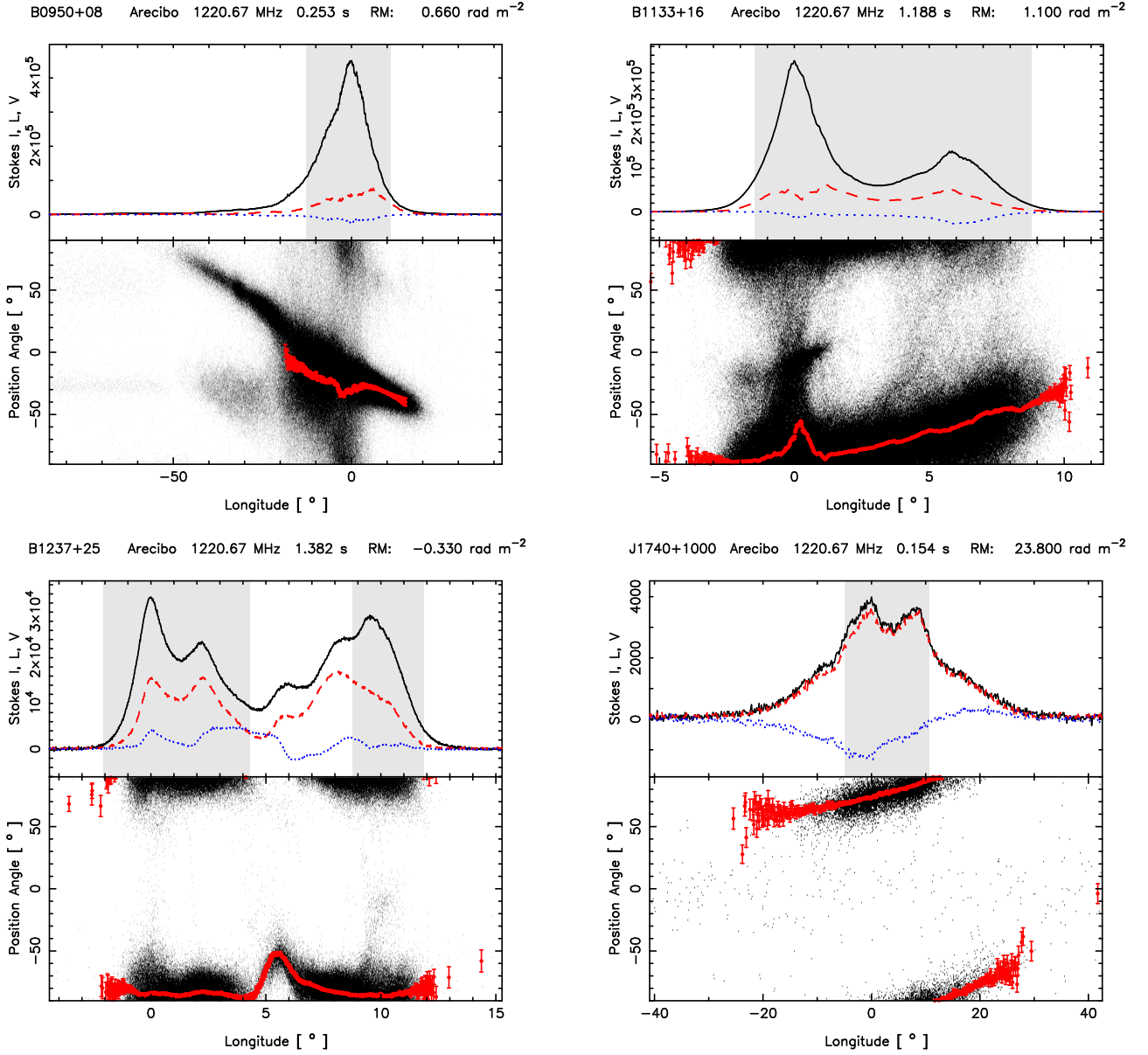

\begin{center}
\begin{tabular}{@{}lr@{}}
 {\mbox{\includegraphics[width=78mm,angle=-90.]{srB0950+08.55522l_a.ps}}}& \ \ \ \ \ \
 {\mbox{\includegraphics[width=78mm,angle=-90.]{srB1133+16.55522l_a.ps}}}\\
 {\mbox{\includegraphics[width=78mm,angle=-90.]{srB1237+25.56550l_a.ps}}}& \ \ \ \ \ \
 {\mbox{\includegraphics[width=78mm,angle=-90.]{srJ1740+1000.55632l_a.ps}}}\\
\end{tabular}
\caption{PPA histograms as in Fig.~\ref{figA1} for pulsars B0950+08, B1133+16, B1237+25 and J1740+1000
observed with time resolutions of 59.5$\mu$sec.}
\label{figA3}
\end{center}
\end{figure*}

\begin{figure*}
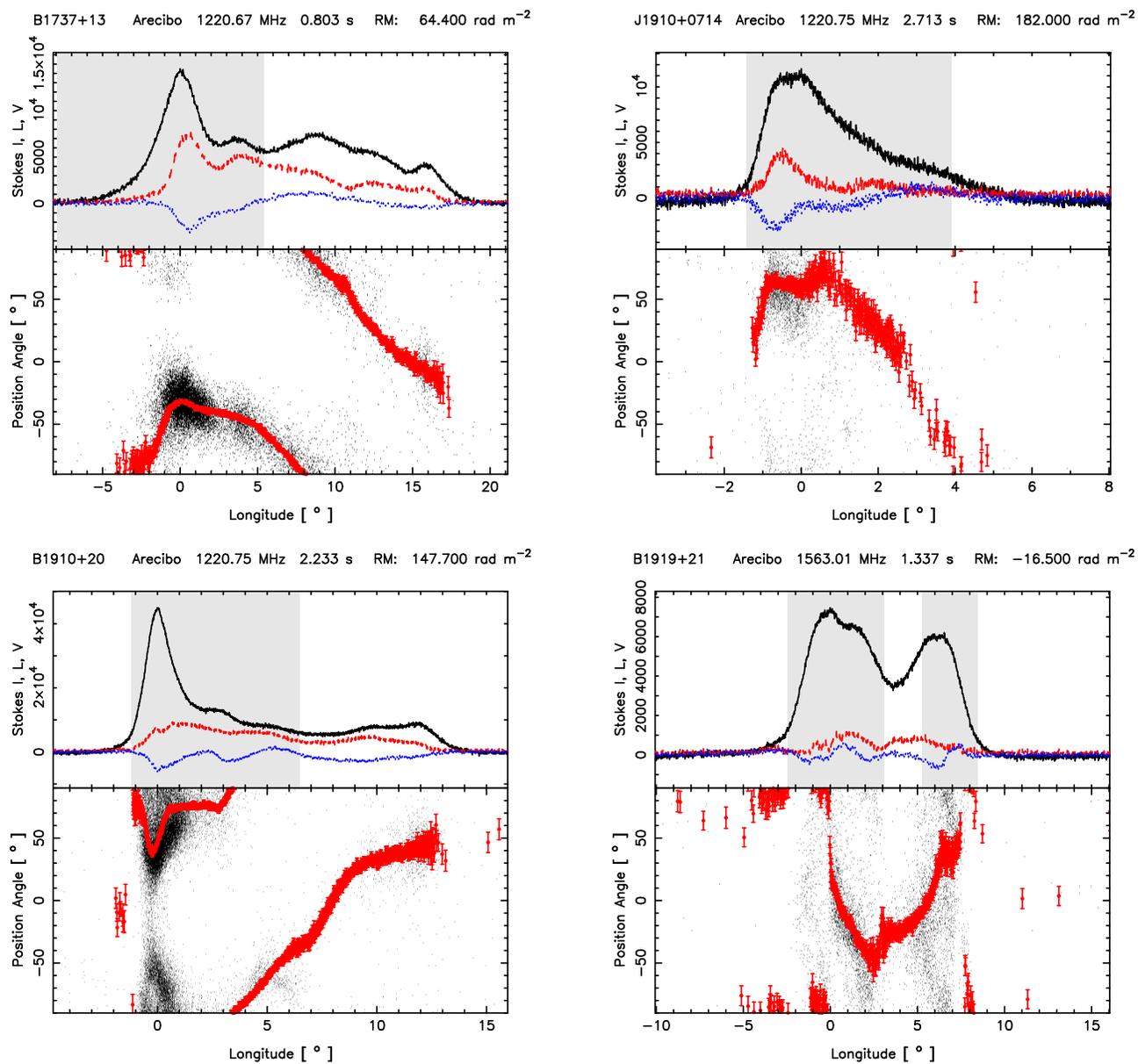

\begin{center}
\begin{tabular}{@{}lr@{}}
 {\mbox{\includegraphics[width=78mm,angle=-90.]{srB1737+13.55632l_a.ps}}}& \ \ \ \ \ \
 {\mbox{\includegraphics[width=78mm,angle=-90.]{srJ1910+0714.56564l_a.ps}}}\\
 {\mbox{\includegraphics[width=78mm,angle=-90.]{srB1910+20.56564l_a.ps}}}& \ \ \ \ \ \
 {\mbox{\includegraphics[width=78mm,angle=-90.]{srB1919+21.56207l_a.ps}}}\\
\end{tabular}
\caption{PPA histograms as in Fig.~\ref{figA1} for pulsars B1737+13, J1910+0714, B1910+20 and B1919+21
observed with time resolutions of 59.5$\mu$sec.}
\label{figA4}
\end{center}
\end{figure*}

\begin{figure*}
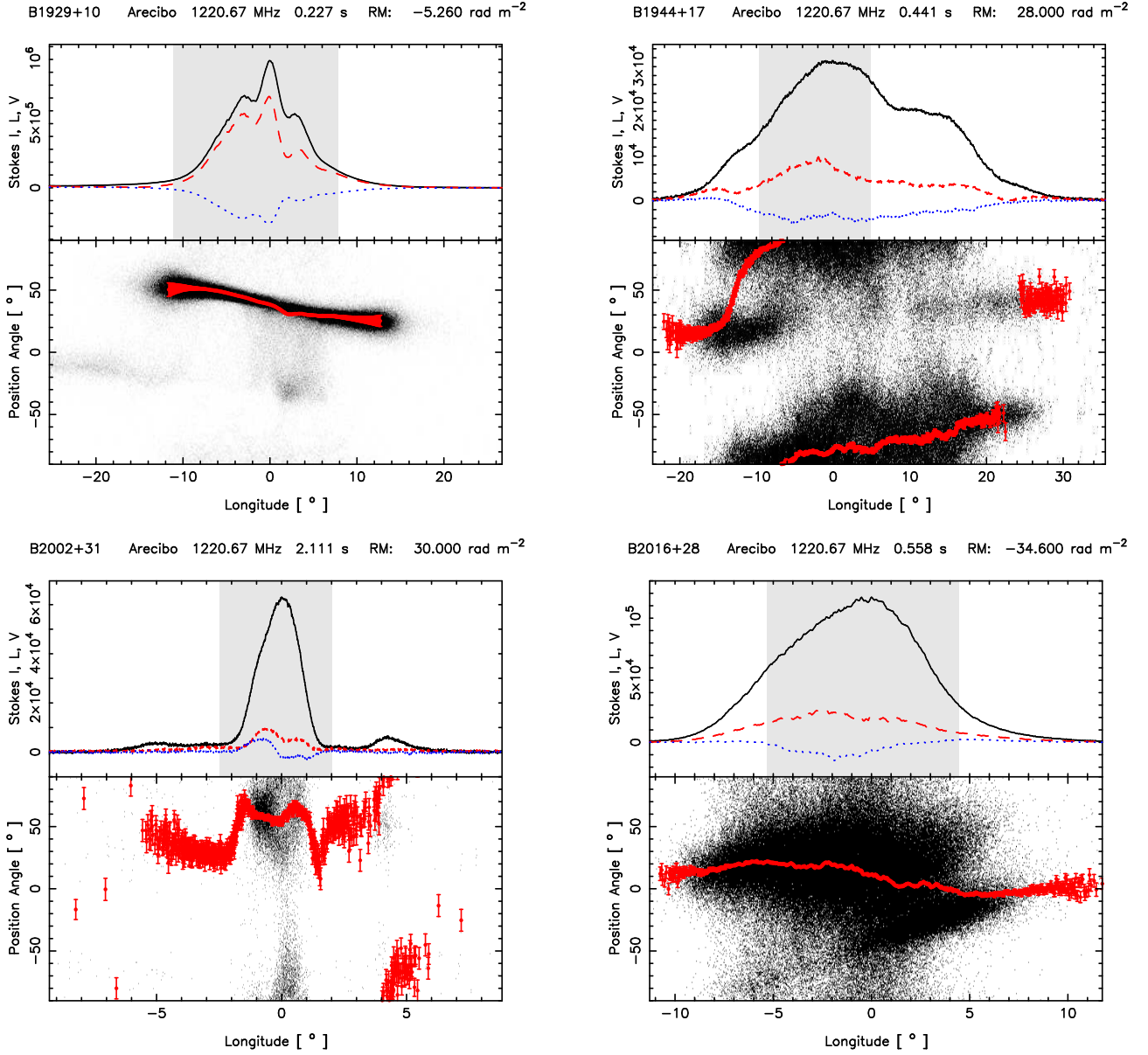

\begin{center}
\begin{tabular}{@{}lr@{}}
 {\mbox{\includegraphics[width=78mm,angle=-90.]{srB1929+10.55633l_a.ps}}}& \ \ \ \ \ \
 {\mbox{\includegraphics[width=78mm,angle=-90.]{srB1944+17.55633l_a.ps}}}\\
 {\mbox{\includegraphics[width=78mm,angle=-90.]{srB2002+31.56564l_a.ps}}}&\ \ \ \ \ \ 
 {\mbox{\includegraphics[width=78mm,angle=-90.]{srB2016+28.55632l_a.ps}}}\\
\end{tabular}
\caption{PPA histograms as in Fig.~\ref{figA1} for pulsars B1929+10, B1944+17, B2002+31 and B2016+28
observed with time resolutions of 59.5$\mu$sec.}
\label{figA5}
\end{center}
\end{figure*}

\begin{figure*}
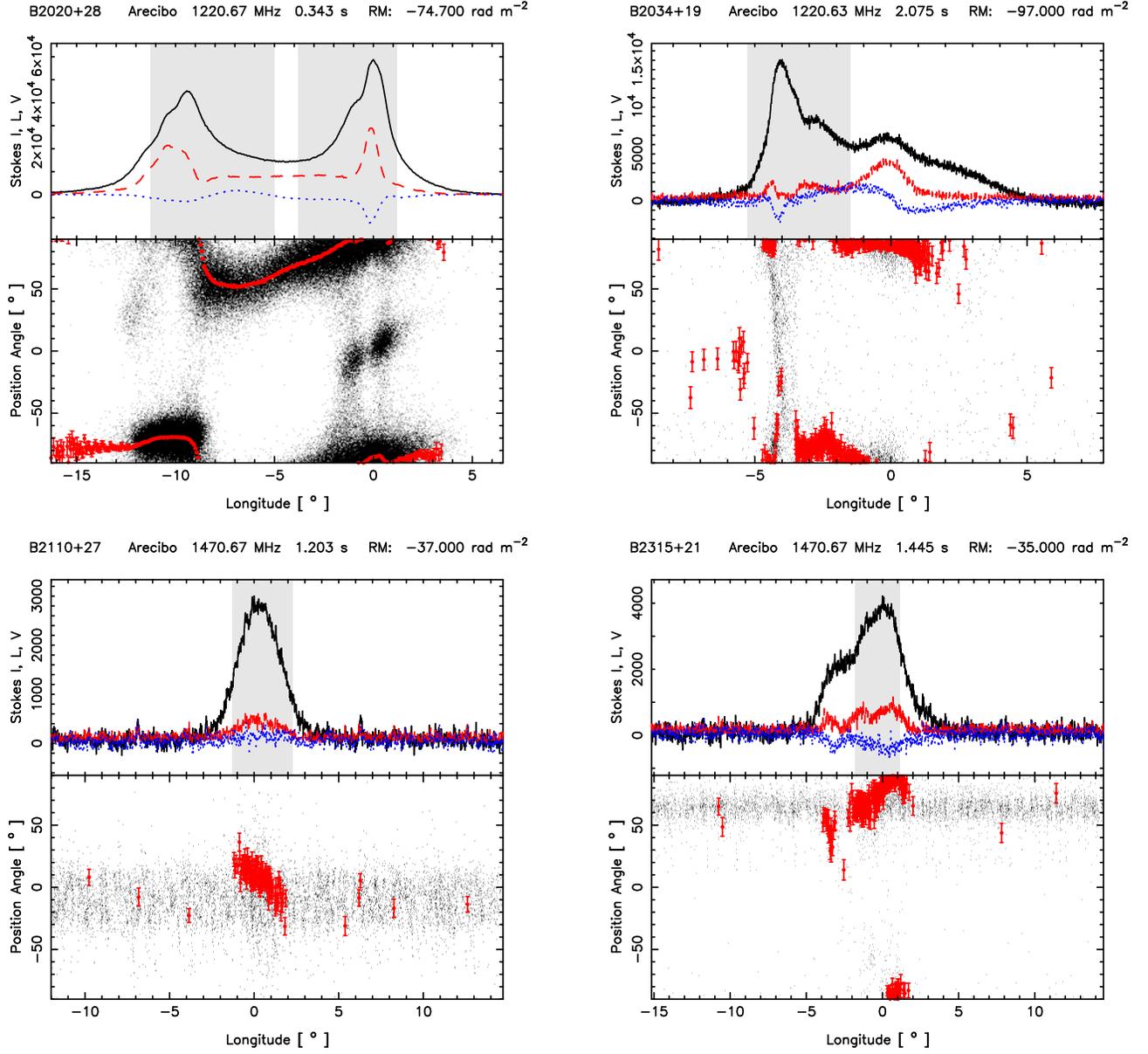

\begin{center}
\begin{tabular}{@{}lr@{}}
 {\mbox{\includegraphics[width=78mm,angle=-90.]{srB2020+28.55632l_a.ps}}}& \ \ \ \ \ \
 {\mbox{\includegraphics[width=78mm,angle=-90.]{srB2034+19.56564l_a.ps}}}\\
 {\mbox{\includegraphics[width=78mm,angle=-90.]{srB2110+27.55425l_a.ps}}}&\ \ \ \ \ \
 {\mbox{\includegraphics[width=78mm,angle=-90.]{srB2315+21.55425l_a.ps}}}\\
\\
\end{tabular}
\caption{PPA histograms as in Fig.~\ref{figA1} for pulsars B2020+28, B2034+19, B2110+27, and B2315+21
observed with time resolutions of 59.5$\mu$sec.}
\label{figA6}
\end{center}
\end{figure*}

\pagebreak
\begin{center}
\textbf{\large Supplemental Materials: Polarized quasiperiodic structures in pulsar radio emission reflect temporal modulations of non-stationary plasma flow}
\end{center}
\setcounter{equation}{0}
\setcounter{figure}{0}
\setcounter{table}{0}
\setcounter{page}{1}
\makeatletter
\renewcommand{\theequation}{S\arabic{equation}}
\renewcommand{\thefigure}{S\arabic{figure}}
\renewcommand{\thetable}{S\arabic{table}}
\renewcommand{\bibnumfmt}[1]{[S#1]}
\section{Results of pulsar microstructure analysis}
\label{a_mres}

The results of our pulsar microstructure timescale analysis pipeline
(as described in Sec.\ 3 and appendix of the main paper)
are available for download as a tarball with the URL\\
\centerline{\texttt{ftp://wm.ncra.tifr.res.in/dmitra/hires\_analysis\_pdf.tar.gz}} 

 or 

\centerline{\texttt{ http://cms.unipune.ac.in/reports/tr-20150223/}}
Untarring this tarball creates the directory \texttt{hires\_analysis\_pdf/}.
This directory contains two pdf files for each pulsar data set analyzed,
with names of the form\\
\centerline{\texttt{<dataset>-<envelope\_smoothing\_bandwidth>-pulses3.pdf}}\\
containing pulsewise fit, envelope, microstructure, and ACF plots
(where rejected pulses are indicated using gray shading overlaid on plots),
and\\
\centerline{\texttt{<dataset>-<envelope\_smoothing\_bandwidth>-hist3.pdf}}\\
containing microstructure timescale histograms.
For example, the file {\small \texttt{B0301+19LC-0p05-pulses3.pdf}}
and {\small \texttt{B0301+19LC-0p05-hist3.pdf}}
respectively contain pulsewise plots and microstructure timescale histograms
for the data set B0301+19LC when analyzed using envelope smoothing bandwidth $h=0.05$.

\begin{deluxetable}{llccccc}
\tablecolumns{7}
\tabletypesize{\footnotesize}
\tablewidth{0pc}
\tablecaption{\label{tabC1}Additional details on the data sets analyzed (compare with Table 2, main paper). Column 7 is the pulse selection threshold on the fit degrees of freedom (see Sec.\ 3, main paper).}
\tablehead{
\colhead{}&\colhead{}&\colhead{}&\colhead{}&\colhead{}&\colhead{}&\colhead{}{} \\
\bf{No.}&{\bf Dataset } &  \bf {Longitude}       & \bf{\# of Pulses in} & \bf{NPT}  & \bf{Outliers} & \bf{Percentile} \\
        &               &  \bf{Range} ($^\circ$) &     \bf{the Dataset} &           & \bf{Curated?} & \bf{Cut-Off}}
\startdata
1. &B0301+19  & -2.2 --- 2.5  &    120      &  301       &    N             &   10 \\
2. &B0301+19LC& -11.8 --- -5.2&    129      &  427       &    N             &   10 \\
3. &J0546+2441&-2.13 --- 1.81 &    105      &  504       &    N             &   10 \\
4. &B0525+21  &-2.85 --- 2.86 &    300      & 1001       &    Y             &   25 \\
5. &B0525+21LC&-14.3 --- -10.9&    300      & 1201       &    Y             &   25\\
6. &B0656+14  &-0.61 --- 4.95 &    155      &  101       &    N             &   5\\
7. &B0751+32  &-2.24 --- 2.43 &     84      &  304       &    N             &   10 \\
8. &B0823+26  &-1.65 --- 1.41 &    226      &  264       &    N             &   10 \\
9. &B0834+06LC&-1.32 --- 1.46 &    100      &  167       &    N             &   10 \\
10. &B0834+06C2&3.7 --- 7.2    &    100      &  210       &    N             &  10\\
11. &B0919+06 &-7.5 --- 4.74  &    104      &  249       &    N             &  10\\
12.&B0950+08  &-12.77 --- 10.84&   102      &  279       &    N             &  10\\
13.&B1133+16LC&-1.47 --- 3.39&     210      &  271       &    N             &  10\\
14.&B1133+16T &3.39   --- 8.8&     210      &  301       &    N             & 25\\
15.&B1237+25 &-2.07 --- 4.3  &     108      &  401       &    N             & 10\\
16.&B1237+25TC&8.77 --- 11.87&     108      &  201       &    N             & 10\\
17.&B1737+13 & -7.94 --- 5.41&      93      &  501       &    N             & 10\\
18.&J1740+1000&-4.87 --- 10.7&     116      &  113       &    N             & 10\\
19.&B1910+20  &-1.21 --- 6.46&     341      &  801       &    N             & 10\\
20.&J1910+0714&-1.41 --- 3.91&     102      & 675        &    N             &10\\
21.&B1919+21LC&-2.45 --- 3.10&      85      &  348       &    N             & 10\\
22.&B1919+21TC&5.28 --- 8.48 &      68      &  201       &    N             & 10\\
23.&B1929+10  &-11.1 --- 7.84&     529      &  201       &    N             & 10\\
24.&B1944+17  &-9.67 --- 4.91&     136      &  301       &    N             & 10\\
25.&B2002+31   &-2.48 --- 2.01&    105      &  445       &    N             & 10\\
26.&B2016+28   &-5.34 --- 4.45&    341      &  256       &    N             & 10\\
27.&B2020+28LC &-11.27 --- -5.03&  348      &  101       &    N             & 10\\
28.&B2020+28   &-3.78 --- 1.21&    580      &   98       &    N             & 10\\
29.&B2034+19   &-5.25 --- -1.49&   134      &  361       &    N             & 10\\
30.&B2110+27   &-1.3 --- 2.26  &   108      &  201       &    N             & 10\\
31.&B2315+21   &-1.79 --- 1.17 &   74       &  201       &    N             & 10
\enddata
\end{deluxetable}

\section{ Ascii Profiles}
\label{asc_prof}
The average profiles obtained for our microstructure observations can be downloaded as 
ascii profiles from \centerline{\texttt{ftp://wm.ncra.tifr.res.in/dmitra/hires\_ascprof.tar.gz}} 

or 

\centerline{\texttt{http://cms.unipune.ac.in/reports/tr-20150223/}} as a tarball.
Untarring the tarball creates a directory ``asc\_profiles" which has the ascii files for
each pulsar. A README file describes the contents of the files.

\section{Plots for average profiles}
\label{c_avprof}
The average profiles and polarization position angle (PPA) histograms
for the pulsars which were not used for microstructure analysis given as non-boldface pulsars 
in Table.~\ref{tab1} are presented.

\begin{figure*}
\begin{center}
\begin{tabular}{@{}lr@{}}
 {\mbox{\includegraphics[width=78mm,angle=-90.]{srB0301+19.55283p1_a.ps}}}& \ \ \ \ \ \
 {\mbox{\includegraphics[width=78mm,angle=-90.]{srB0525+21.56572p_a.ps}}}\\
 {\mbox{\includegraphics[width=78mm,angle=-90.]{srJ0546+2441.56572p_a.ps}}}& \ \ \ \ \ \
 {\mbox{\includegraphics[width=78mm,angle=-90.]{srB1929+10.55320p1_a.ps}}}\\
\end{tabular}
\caption{PPA histograms for pulsars PSR B0301+19, B0525+21, J0546+2441 and B1929+10 observed with time
resolution of 59.5 $\mu$sec, 
where the instrument and band is indicated above each plot. The respective
upper panels give the total power (black), total linear (red) and circular polarization LH-RH (blue).
The lower panels give the polarization-angle density as black dots and the 
average PPA is plotted as red points.}
\label{figC1}
\end{center}
\end{figure*}

\begin{figure*}
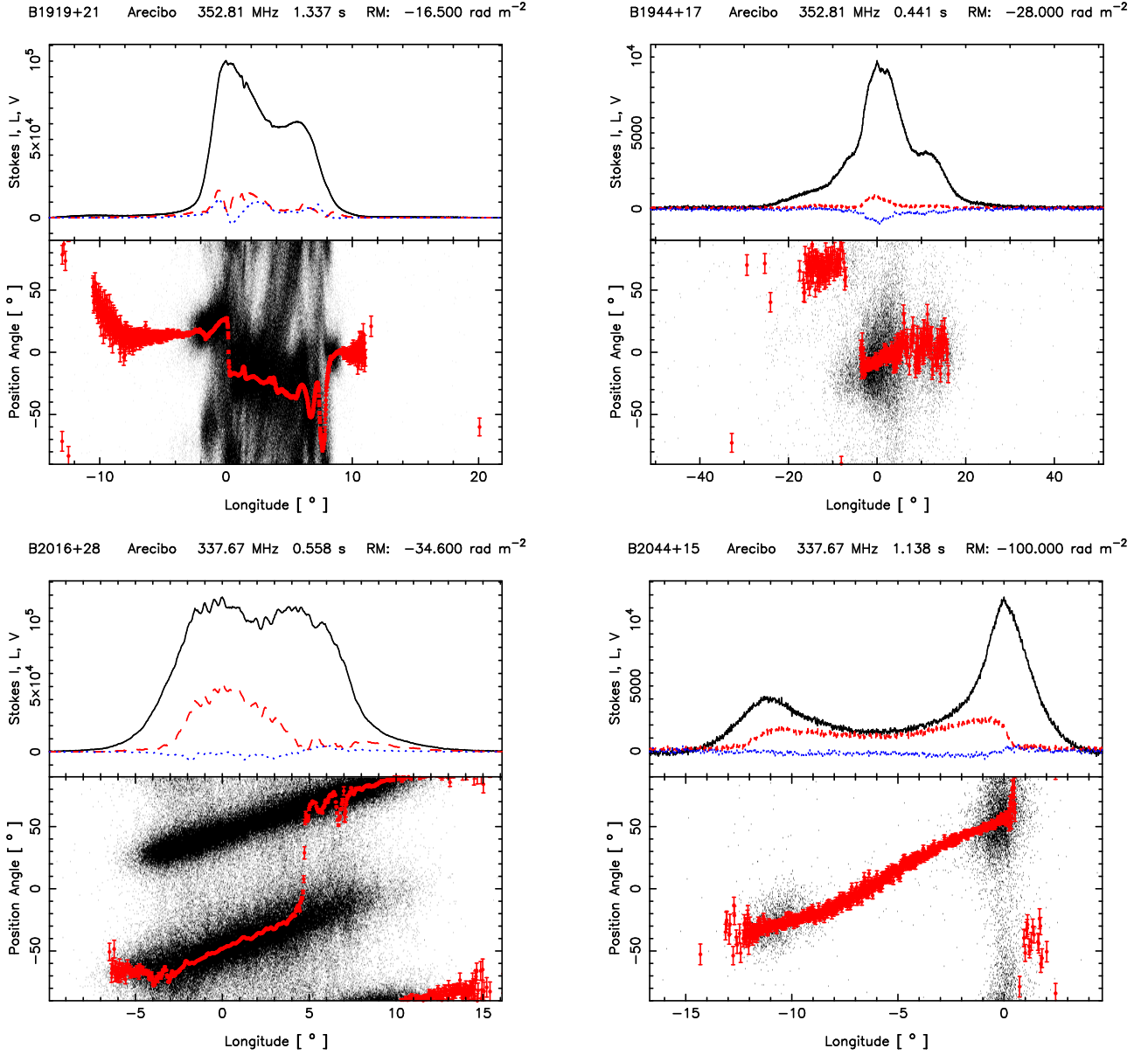

\begin{center}
\begin{tabular}{@{}lr@{}}
 {\mbox{\includegraphics[width=78mm,angle=-90.]{srB1919+21.55320p_a.ps}}}& \ \ \ \ \ \
 {\mbox{\includegraphics[width=78mm,angle=-90.]{srB1944+17.55320p_a.ps}}}\\
 {\mbox{\includegraphics[width=78mm,angle=-90.]{srB2016+28.55320p1_a.ps}}}& \ \ \ \ \ \
 {\mbox{\includegraphics[width=78mm,angle=-90.]{srB2044+15.55320p1_a.ps}}}\\
\end{tabular}
\caption{PPA histograms as in Fig.~\ref{figC1} for pulsars B1919+21, B1944+17, B2016+28 
and B2044+15 observed with time resolutions of 59.5 $\mu$sec.}
\label{figC2}
\end{center}
\end{figure*}

\begin{figure*}
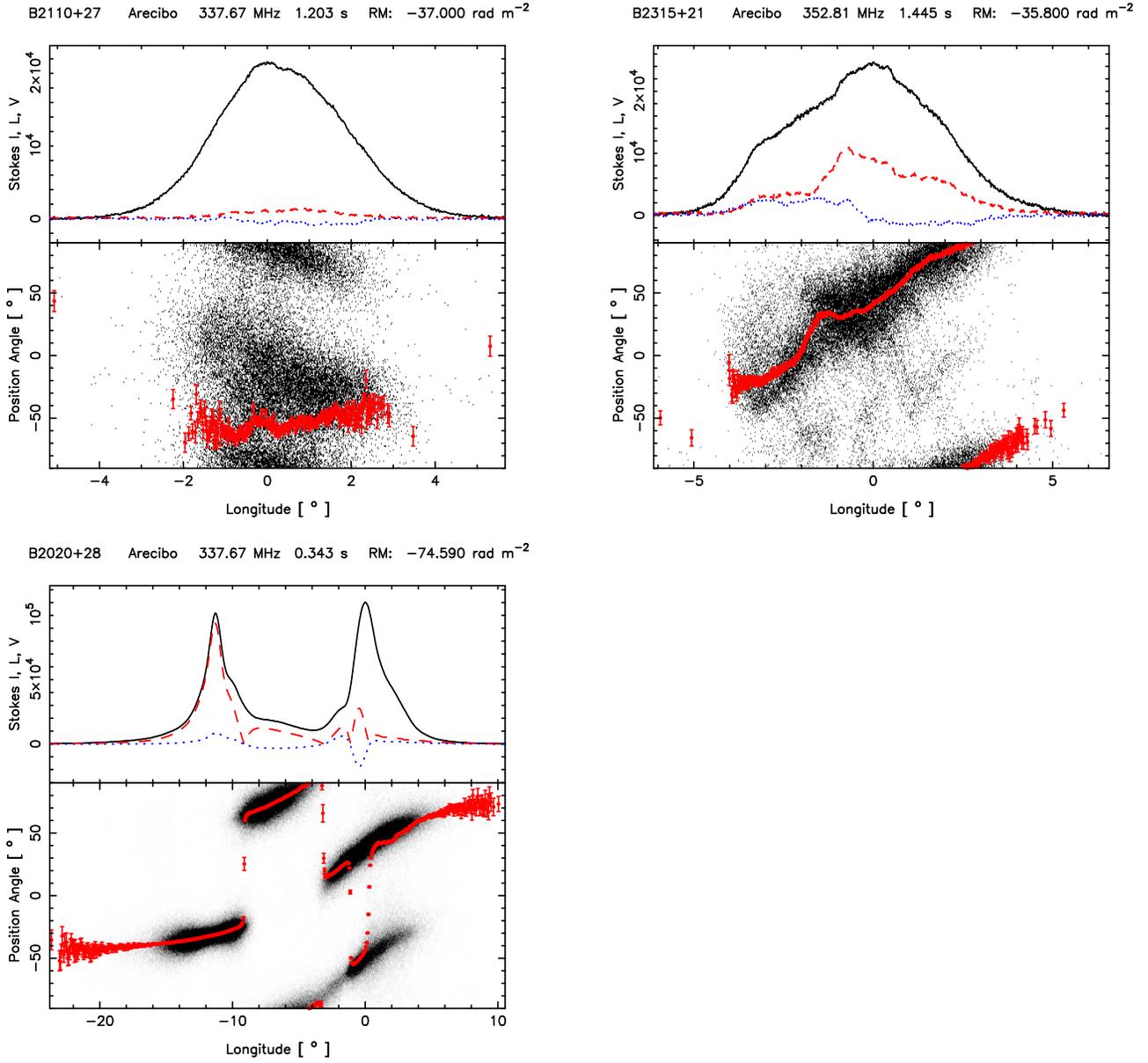

\begin{center}
\begin{tabular}{@{}lr@{}}
 {\mbox{\includegraphics[width=78mm,angle=-90.]{srB2110+27.55321p1_a.ps}}}& \ \ \ \ \ \
 {\mbox{\includegraphics[width=78mm,angle=-90.]{srB2315+21.55275p_a.ps}}}\\
 {\mbox{\includegraphics[width=78mm,angle=-90.]{srB2020+28.55321p1_a.ps}}}& \ \ \ \ \ \
\\
\end{tabular}
\caption{PPA histograms as in Fig.~\ref{figC1} for pulsars B2110+27, B2315+21 and B2020+28 
observed with time resolutions of 59.5 $\mu$sec.}
\label{figC3}
\end{center}
\end{figure*}
\begin{figure*}
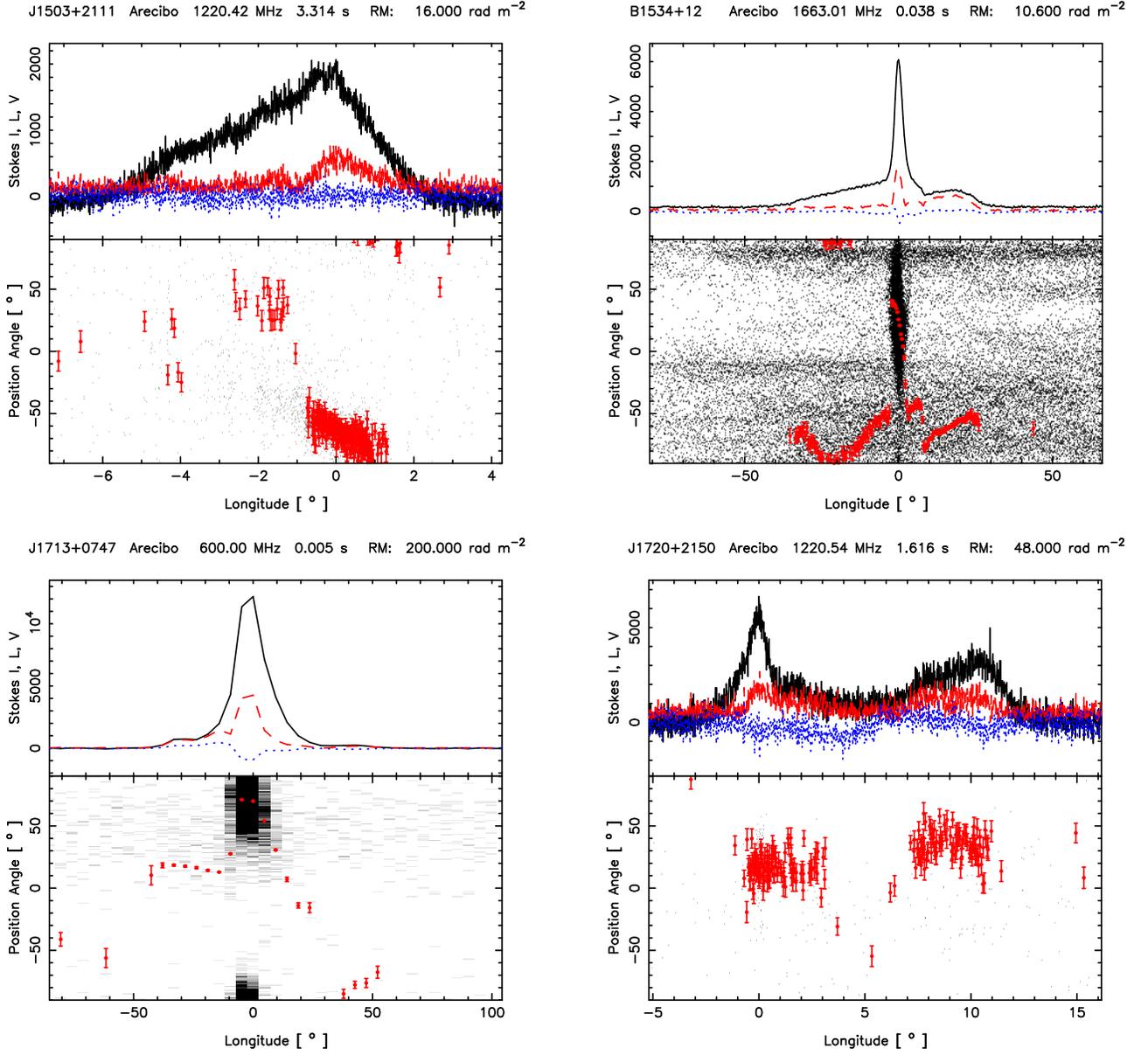

\begin{center}
\begin{tabular}{@{}lr@{}}
 {\mbox{\includegraphics[width=78mm,angle=-90.]{srJ1503+2111.56550l_a.ps}}}& \ \ \ \ \ \
 {\mbox{\includegraphics[width=78mm,angle=-90.]{srB1534+12.55637l_a.ps}}}\\
 {\mbox{\includegraphics[width=78mm,angle=-90.]{srJ1713+0747.55632l_a.ps}}}& \ \ \ \ \ \
 {\mbox{\includegraphics[width=78mm,angle=-90.]{srJ1720+2150.56550l_a.ps}}}\\
\\
\end{tabular}
\caption{PPA histograms as in Fig.~\ref{figC1} for pulsars J1503+2111, B1534+12, J1713+0747 and J1720+2150
observed with time resolutions of 59.5 $\mu$sec.}
\label{figC4}
\end{center}
\end{figure*}

\begin{figure*}
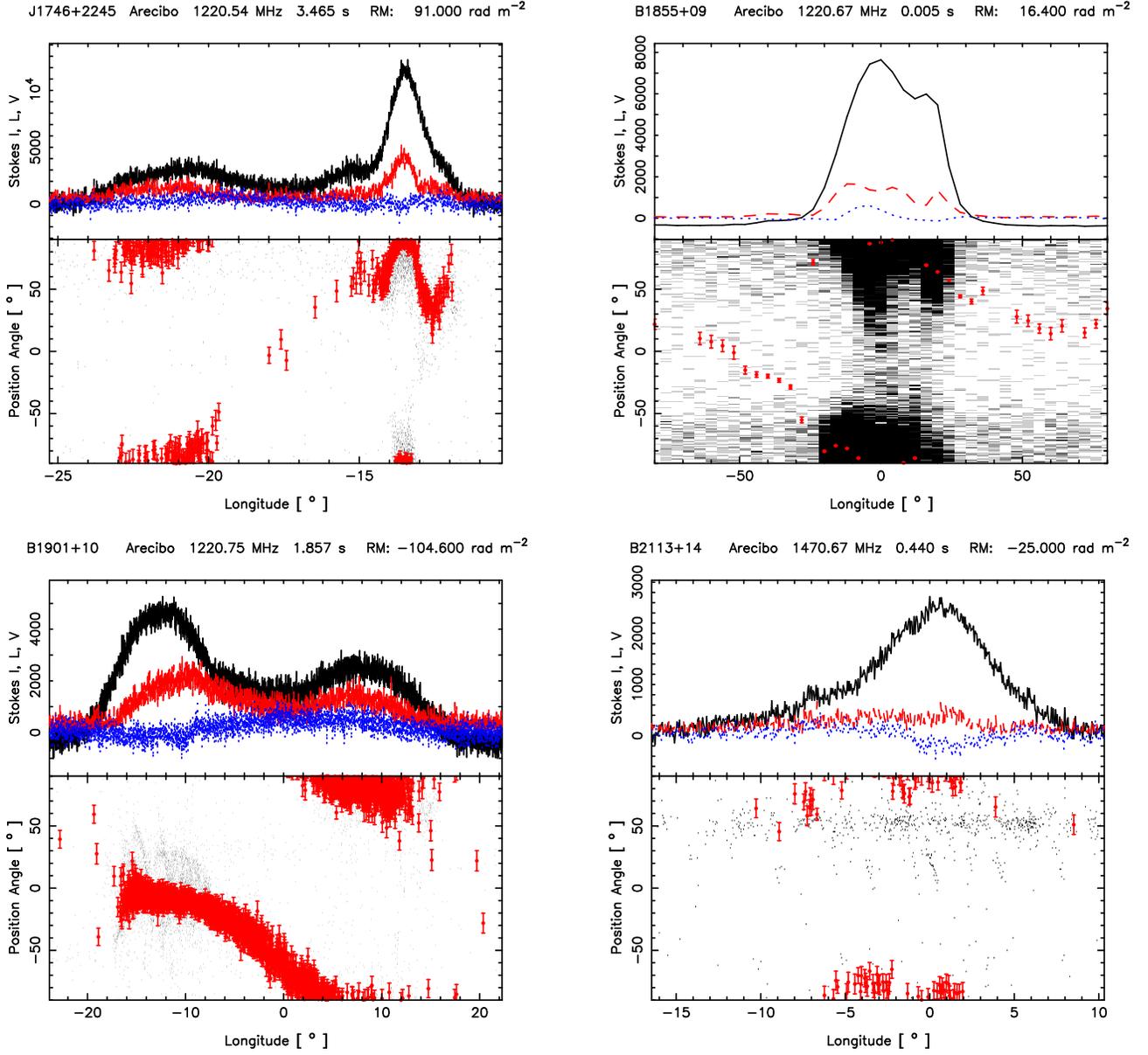

\begin{center}
\begin{tabular}{@{}lr@{}}
 {\mbox{\includegraphics[width=78mm,angle=-90.]{srJ1746+2245.56550l_a.ps}}}& \ \ \ \ \ \
 {\mbox{\includegraphics[width=78mm,angle=-90.]{srB1855+09.55637l_a.ps}}}\\
 {\mbox{\includegraphics[width=78mm,angle=-90.]{srB1901+10.56564l_a.ps}}}& \ \ \ \ \ \
 {\mbox{\includegraphics[width=78mm,angle=-90.]{srB2113+14.55425l_a.ps}}}\\
\\
\end{tabular}
\caption{PPA histograms as in Fig.~\ref{figC1} for pulsars J1746+2245, B1855+09, B1901+10 and B2113+14 
observed with time resolutions of 59.5 $\mu$sec.}
\label{figC5}
\end{center}
\end{figure*}

\end{document}